\documentclass[twocolumn,twocolappendix]{aastex631}

\usepackage{amsmath}
\usepackage{amssymb}
\usepackage{amsfonts}
\usepackage{amsthm}
\usepackage{soul}

\hypersetup{linkcolor=cyan,citecolor=cyan,filecolor=cyan,urlcolor=cyan}

\begin{document}

\title{Jet interaction with galaxy cluster mergers}

\author[0000-0001-7058-8418]{P. Dom\'inguez-Fern\'andez}
\affiliation{Harvard-Smithsonian Center for Astrophysics, 60 Garden Street, Cambridge, MA 02138, USA}
\correspondingauthor{Paola Dom\'inguez-Fern\'andez}
\email{pdominguezfernandez@cfa.harvard.edu}

\author[0000-0003-3175-2347]{J. ZuHone}
\affiliation{Harvard-Smithsonian Center for Astrophysics, 60 Garden Street, Cambridge, MA 02138, USA}

\author[0000-0001-6260-9709]{R. Weinberger}
\affiliation{Leibniz-Institut f\"ur Astrophysik Potsdam (AIP), An der Sternwarte 16, 14482 Potsdam, Germany}

\author[0000-0001-6411-3686]{E. Bellomi}
\affiliation{Harvard-Smithsonian Center for Astrophysics, 60 Garden Street, Cambridge, MA 02138, USA}

\author[0000-0001-6950-1629]{L. Hernquist}
\affiliation{Harvard-Smithsonian Center for Astrophysics, 60 Garden Street, Cambridge, MA 02138, USA}

\author[0000-0003-0297-4493]{P. Nulsen}
\affiliation{Harvard-Smithsonian Center for Astrophysics, 60 Garden Street, Cambridge, MA 02138, USA}

\author[0000-0003-4195-8613]{G. Brunetti}
\affiliation{1INAF/Osservatorio di Radioastronomia, via Gobetti 101, I-40129 Bologna, Italy}




\begin{abstract}

AGN bubbles in cool-core galaxy clusters are believed to  
facilitate the transport of cosmic ray electrons (CRe) throughout the cluster. Recent radio observations are revealing complex morphologies of cluster diffuse emission, potentially linked to interactions between AGN bursts and the cluster environment. We perform three-dimensional magneto-hydrodynamical simulations of binary cluster mergers and inject a bi-directional jet at the center of the main cluster. Kinetic, thermal, magnetic and CR energy are included in the jet and we use the two-fluid formalism to model the CR component. We explore a wide range of cluster merger and jet parameters. We discuss the formation of various wide-angle-tail (WAT) and X-shaped sources in the early evolution of the jet and merger. During the last phase of the evolution, we find that the CR material efficiently permeates the central region of the cluster reaching radii of $\sim1$--2 Mpc within $\sim5$--6 Gyr, depending on the merger mass ratio. We find that solenoidal turbulence dominates during the binary merger and explore the possibility for the CR jet material to be re-accelerated by super-Alfv\`enic turbulence and contribute to cluster scale radio emission. 
We find high volume fractions, $\gtrsim 70$\%, at which the turbulent acceleration time is shorter than the electron cooling time.
Finally, we study the merger shock interaction with the CRe material and show that it is unlikely that this material significantly contributes to the radio relic emission
associated with the shocks. We suggest that multiple jet outbursts and/or off-center radio galaxies would increase the likelihood of detecting these merger shocks in the radio due to shock re-acceleration.
\end{abstract}

\keywords{Galaxy clusters (584) --- Intracluster medium (858) --- Magnetohydrodynamics (1964) --- High energy astrophysics (739) --- Shocks (2086) --- Plasma astrophysics (1261)}


\section{Introduction} \label{sec:intro}

The diffuse synchrotron radio emission observed in galaxy clusters is typically produced by cosmic-ray (CR) electrons, with Lorentz factors ranging from $\gamma\sim10^3$ to $10^4$, emitting in microgauss magnetic fields \citep[][]{2012A&ARv..20...54F,2019SSRv..215...16V}. 
The origin of these CR electrons (CRe) remains uncertain. 
Thermal electrons can be energized to CR energies through different acceleration mechanisms and potentially explain the synchrotron emission. Yet, in the majority of cases, the energy of these thermal electrons is insufficient to account for the observations \citep[see e.g.][for the case of radio relics]{2020A&A...634A..64B}.
Therefore, sources of mildly relativistic electrons have frequently been invoked as necessary to explain some of the diffuse radio emission in radio relics through diffusive shock acceleration \citep[DSA;][]{dv81, 2016ApJ...823...13K}; giant radio haloes \citep[see][]{2019SSRv..215...16V} and the radio emission observed in some cool-core clusters \citep[e.g.,][]{2014MNRAS.444L..44B, 2024arXiv240309802B} through turbulent re-acceleration \citep[][]{2001MNRAS.320..365B,2001ApJ...557..560P}. This population of fossil CRe could be injected by jets launched by AGN \citep[][]{2012ARA&A..50..455F} and/or by hadronic processes \citep[secondary electrons; see][]{1999APh....12..169B,2004A&A...413...17P,2005MNRAS.363.1173B} and/or 
by shocks \citep[][]{2003ApJ...593..599R}, merger turbulence \citep[][]{2005MNRAS.364..753D,2009A&A...504...33V} and/or sloshing motions \citep[][]{2008ApJ...675L...9M,2013ApJ...762...78Z} in the intracluster medium (ICM; see \citealt{2014IJMPD..2330007B} for a review).


The often reported tailed and bent-tailed radio galaxies embedded within the ICM supply CRe to the cluster environment. The evolution of this remnant AGN plasma can be further affected by the dynamics of the galaxy cluster. For instance, gas sloshing motions and mild shocks can distort AGN radio lobes, dispersing their relativistic material up to several hundred kpc or even Mpc from their initial location. \citep[][]{2021ApJ...914...73Z,2021A&A...653A..23V,2022MNRAS.510.4000F}. Shock waves can sweep through a radio bubble,  compressing it adiabatically and re-energizing the fossil plasma \citep[][]{2001A&A...366...26E}. These sources, referred to as ``phoenices'' in the literature, are mostly irregular sources with a steep spectrum and without clear spatial trends. Another example are the so-called gently re-energized tails \citep[GRaETs][]{2017SciA....3E1634D} of radio galaxies with spectral aging starting from the host radio galaxy and with a sudden re-brightening at the end of the tails coinciding with a flatter spectral index \citep[see e.g.,][]{2022A&A...666A...3E,2022A&A...659A..20I}. Finally, this AGN fossil plasma is also thought to be essential for the formation of mini-halos \citep[][]{2020MNRAS.499.2934R}. With the detection of an increasing number of these sources, our comprehension of the widespread presence of fossil plasma pools in clusters deepens, highlighting the need for more extensive theoretical and numerical investigations.

In addition to CRe, it is believed that the inflation of radio lobes and expansion of radio jets contribute to the injection of energy into the ICM, thereby playing a significant role in controlling its thermodynamic properties \citep[e.g.,][]{2004ApJ...607..800B,2011ApJ...735...11O,2021Univ....7..142E,2022PhR...973....1D}. Central AGN activity is capable of channeling mechanical energy through radio jets, which, as they inflate cavities, work against the ambient gas pressure as observed in X-rays \citep[][]{2000ApJ...534L.135M}. While the conditions that lead to the onset of AGN activity are not fully known, it is clear that multiple cycles of activity can occur as hinted by the existence of multiple X-ray cavities due to the radio activity in the brightest central galaxy \citep[][]{2005MNRAS.360L..20F,2014MNRAS.442.3192V,2015ApJ...805...35H,2024ApJ...961..134U}. These succesive phases of radio activity can result in complex radio morphologies of the injected plasma, indicating different levels of mixing with the ICM \citep[e.g.,][]{2018MNRAS.473.3536W,2020A&A...634A...4M}.


In \cite{2021Galax...9...91Z,2021ApJ...914...73Z}, the authors proposed that AGN bubbles could evolve into radio relics. In particular, they suggested that merger-driven gas motions could transport CRe to larger radii in the tangential direction. The observed Mpc-scale thin filaments of jet material, coupled with the amplification of magnetic fields parallel to these structures, suggest the possibility that such dynamics could account for the appearance of radio relics.
In the present paper, we generalize this idea by doing a parameter study of different binary merger setups while a central black hole injects jets into the ICM. 
The focus of this work is to study the spatial distribution of CRs injected by a single AGN burst during a binary merger event. 
Our first aim is to explore how plausible it is for binary merger dynamics to re-accelerate the fossil material either through the merger shocks or the available turbulence.
Our second aim is to understand under what conditions the CRs could permeate a large central region of a galaxy cluster. A follow-up paper will include the synchrotron modelling for our simulations.

The paper is structured as follows. In Section \ref{section:num_set-up}, we describe our numerical set-up and initial conditions. In Section \ref{sec:results} we show our results. We discuss our results in the context of observed radio galaxies, radio relics and radio halos in sub-sections \ref{sec:sources}--\ref{sec:halos}. We summarize our work in Section \ref{sec:conclusions}.


\section{Numerical set-up}\label{section:num_set-up}

\subsection{Initial conditions}
\label{sec:profiles}

We carry out simulations of idealized binary mergers of
two spherically symmetric galaxy clusters in hydrostatic and virial equilibrium following the method described in \cite{2011ApJ...728...54Z}. The gas in each cluster is fully ionized, with an assumed mean molecular weight $\mu = 0.6$, ideal, with a constant adiabatic index $\gamma_0 = 5/3$, and, magnetized, with an initial plasma beta $\beta_p = P_{th}/P_B = 100$. We follow the procedure
of \cite{2019ApJ...883..118B} to set up a divergence-free Gaussian random magnetic field with a Kolmogorov spectrum with exponential cutoffs at wavelengths $k_0=2\pi / \lambda_0$ and $k_1=2\pi / \lambda_1$. The corresponding scales are $\lambda_0=10$ kpc and $\lambda_0=500$ kpc.

We explored mass ratios $R=M_2$:$M_1$ of 1:2 and 1:5,
in which the primary cluster has an initial mass of $M_1 = M_{200,1}= 5.9 \times 10^{14} M_{\odot}$\footnote{$M_{200}$ is the mass enclosed within a radius $r_{200}$ where $r_{200}$ is the radius at which the mean density of the cluster is 200 times the critical density.}. This main cluster is assumed to be a Perseus-like, relaxed, cool-core cluster. This assumption is appropriate because central radio AGN in galaxy clusters are predominantly found in cool-core environments rather than non-cool-core environments \citep[see e.g.][]{2018MNRAS.480L..68I}. In future work, we will also pursue other initial conditions.

We closely follow the initial set-up from \cite{2023arXiv231009422B} where a super-Navarro-Frenk-White (sNFW) profile \citep[see][]{2018MNRAS.476.2086L} is assumed for the total mass density of the main cluster:
\begin{equation}
    \rho_{\mathrm{sNFW}}(r) = \frac{3M}{16\pi a^3} \frac{1}{x \left( 1 + x \right)^{5/2}},
\end{equation}
where $M=1.5 \times 10^{15} \, M_{\odot}$ is the total mass\footnote{Note that $M$ is the total mass integrated to $r \rightarrow \infty$.} of the main cluster dark matter (DM) halo, $x = r/a$ and $a=389.5$ kpc is its scale radius. The electron number density of the main cluster is modeled by a sum of a $\beta$-model profile \citep[][]{1976A&A....49..137C} and the modified $\beta$-model profile from
\cite{2006ApJ...640..691V}:
\begin{equation}\label{eq:density_profile}
\begin{split}
    &n_e(r) = n_{e,c1} \left[ 1 + \left(  \frac{r}{r_{c1}}
\right)^2 
\right]^{-3 \beta_1/2} + \\
& n_{e,c2} \left[ 1 + \left(  \frac{r}{r_{c2}}
\right)^2 
\right]^{-3 \beta_2/2} \left[ 1 + \left(  \frac{r}{r_{s}}
\right)^{\gamma_s}  \right]^{-\varepsilon/2\gamma_s},
\end{split}
\end{equation}
where $n_{e,c1}=4.5 \times 10^{-2}$ cm$^{-3}$ and $n_{e,c2}=4 \times 10^{-3}$ cm$^{-3}$, $r_{c1}=55$ kpc, $r_{c2}=180$ kpc, $r_s = 1800$ kpc, $\beta_1=1.2$, $\beta_2=0.58$, $\gamma_s=3$ and $\varepsilon=3$. The second cluster has an initial mass of $M_2 = R \, M_1$ and the number density profile of its ICM is given by the second term in Eq.~\ref{eq:density_profile}. In this case, we have $r_s = 1000$ kpc, $\beta_2=2/3$, $\gamma_s=3$ and $\varepsilon=3$. We derived $n_{e,c2}$ by using the scaling relations determined by \cite{2009ApJ...692.1033V} (see Eqs.~8--9). The scale radius is $r_{c2}=222.5$ kpc and $r_{c2}=318.6$ kpc for $R=1:5$ and $R=1:2$, respectively.

Similarly to \cite{2021ApJ...914...73Z} and \cite{2023arXiv231009422B}, for all our simulations, the clusters start at a separation
of $d = 3$ Mpc along the +y-axis from the main cluster center, and with an initial impact parameter $b
\simeq d \sin \theta$. The corresponding initial velocity vector of the second cluster is $\mathbf{\vec{v}_2} = (v_{2,0} \sin \theta , -v_{2,0} \cos \theta, 0)$, where $v_{2,0} \simeq 1370$ km/s. We tried two different angles, $\theta=20^{\circ}$ and $\theta=10^{\circ}$, corresponding to initial impact parameters of $b\simeq 1026$ kpc and $b\simeq 520$ kpc, respectively.

\subsection{AREPO jet simulations}
\label{sec:AREPO}

\begin{table*}
\centering
\begin{tabular}{ccccc}
    & & &&\\ \hline \hline
      ID & R & $\theta$ [$^{\circ}$]  & Jet direction & $t_{\mathrm{jet},0}$ [Gyr] \\ 
\hline
     R0p5\_jetdir-x\_50Myr\_theta20$^{\bigstar}$ &  1:2 & 20  & $x$ & 0.05  \\
     R0p5\_jetdir-y\_50Myr\_theta20 &  1:2 & 20  & $y$ & 0.05  \\
     R0p5\_jetdir-z\_50Myr\_theta20 &  1:2 & 20  & $z$ & 0.05  \\
     R0p5\_jetdir-x\_1Gyr\_theta20 &  1:2 & 20  & $x$ & 1  \\
     R0p5\_jetdir-y\_1Gyr\_theta20 &  1:2 & 20  & $y$ & 1  \\
     R0p5\_jetdir-z\_1Gyr\_theta20 &  1:2 & 20  & $z$ & 1  \\
     R0p5\_jetdir-x\_2Gyr\_theta20 &  1:2 & 20  & $x$ & 2  \\
     R0p5\_jetdir-y\_2Gyr\_theta20 &  1:2 & 20  & $y$ & 2  \\
     R0p5\_jetdir-y\_2Gyr\_theta20 &  1:2 & 20  & $z$ & 2  \\
      R0p5\_jetdir-x\_5Gyr\_theta20 &  1:2 & 20  & $x$ & 5  \\
     R0p5\_jetdir-y\_5Gyr\_theta20 &  1:2 & 20  & $y$ & 5  \\
     R0p5\_jetdir-y\_5Gyr\_theta20 &  1:2 & 20  & $z$ & 5  \\
      \hline
      R0p5\_jetdir-x\_50Myr\_theta10 &  1:2 & 10 & $x$ & 0.05  \\
     R0p5\_jetdir-y\_50Myr\_theta10 &  1:2 & 10  & $y$ & 0.05  \\
     R0p5\_jetdir-z\_50Myr\_theta10 &  1:2 & 10  & $z$ & 0.05  \\
     R0p5\_jetdir-x\_1Gyr\_theta10 &  1:2 & 10  & $x$ & 1  \\
     R0p5\_jetdir-y\_1Gyr\_theta10 &  1:2 & 10  & $y$ & 1  \\
     R0p5\_jetdir-z\_1Gyr\_theta10 &  1:2 & 10  & $z$ & 1  \\
     R0p5\_jetdir-x\_2Gyr\_theta10 &  1:2 & 10  & $x$ & 2  \\
     R0p5\_jetdir-y\_2Gyr\_theta10 &  1:2 & 10  & $y$ & 2  \\
     R0p5\_jetdir-y\_2Gyr\_theta10 &  1:2 & 10  & $z$ & 2  \\
      R0p5\_jetdir-x\_5Gyr\_theta10 &  1:2 & 10  & $x$ & 5  \\
     R0p5\_jetdir-y\_5Gyr\_theta10 &  1:2 & 10  & $y$ & 5  \\
     R0p5\_jetdir-y\_5Gyr\_theta10 &  1:2 & 10  & $z$ & 5  \\
      \hline
      R0p2\_jetdir-x\_50Myr\_theta20$^{\bigstar}$ &  1:5 & 20  & $x$ & 0.05  \\
     R0p2\_jetdir-y\_50Myr\_theta20 &  1:5 & 20 & $y$ & 0.05  \\
     R0p2\_jetdir-z\_50Myr\_theta20 &  1:5 & 20  & $z$ & 0.05  \\
     R0p2\_jetdir-x\_1Gyr\_theta20$^{\bigstar}$ &  1:5 & 20  & $x$ & 1  \\
     R0p2\_jetdir-y\_1Gyr\_theta20 &  1:5 & 20  & $y$ & 1  \\
     R0p2\_jetdir-z\_1Gyr\_theta20 &  1:5 & 20  & $z$ & 1  \\
     R0p2\_jetdir-x\_2Gyr\_theta20$^{\bigstar}$ &  1:5 & 20  & $x$ & 2  \\
     R0p2\_jetdir-y\_2Gyr\_theta20 &  1:5 & 20  & $y$ & 2  \\
     R0p2\_jetdir-y\_2Gyr\_theta20 &  1:5 & 20  & $z$ & 2  \\
     R0p2\_jetdir-x\_5Gyr\_theta20$^{\bigstar}$ &  1:5 & 20  & $x$ & 5  \\
     R0p2\_jetdir-y\_5Gyr\_theta20 &  1:5 & 20  & $y$ & 5  \\
     R0p2\_jetdir-y\_5Gyr\_theta20 &  1:5 & 20  & $z$ & 5  \\
     \hline
     R0p2\_jetdir-x\_50Myr\_theta10 &  1:5 & 10  & $x$ & 0.05  \\
     R0p2\_jetdir-y\_50Myr\_theta10 &  1:5 & 10  & $y$ & 0.05  \\
     R0p2\_jetdir-z\_50Myr\_theta10 &  1:5 & 10 & $z$ & 0.05  \\
     R0p2\_jetdir-x\_1Gyr\_theta10 &  1:5 & 10  & $x$ & 1  \\
     R0p2\_jetdir-y\_1Gyr\_theta10 &  1:5 & 10  & $y$ & 1  \\
     R0p2\_jetdir-z\_1Gyr\_theta10 &  1:5 & 10  & $z$ & 1  \\
     R0p2\_jetdir-x\_2Gyr\_theta10 &  1:5 & 10  & $x$ & 2  \\
     R0p2\_jetdir-y\_2Gyr\_theta10 &  1:5 & 10  & $y$ & 2  \\
     R0p2\_jetdir-y\_2Gyr\_theta10 &  1:5 & 10  & $z$ & 2  \\
     R0p2\_jetdir-x\_5Gyr\_theta10 &  1:5 & 10  & $x$ & 5  \\
     R0p2\_jetdir-y\_5Gyr\_theta10 &  1:5 & 10  & $y$ & 5  \\
     R0p2\_jetdir-y\_5Gyr\_theta10 &  1:5 & 10  & $z$ & 5  \\
     \hline 
\end{tabular}
\caption{Simulation runs. Columns: Simulation ID, merger mass ratio, impact parameter (b=$3000 \sin \theta$ kpc), jet direction, time for jet ignition. The symbol $^{\bigstar}$ indicates those simulations that were also re-simulated at high-resolution.
}
\label{table:sims}
\end{table*}

We carry out 3D non-radiative magnetohydrodynamic (MHD) simulations using the moving-mesh code AREPO \citep[][]{2010MNRAS.401..791S,2011MNRAS.418.1392P} which employs a finite-volume Godunov method on an unstructured moving
Voronoi mesh to solve the ideal MHD equations, and a Tree-PM solver to compute the
self-gravity from gas and dark matter. The MHD Riemann problems at cell interfaces
are solved using a Harten-Lax-van Leer-Discontinuities (HLLD) Riemann solver \citep[see][]{2011MNRAS.418.1392P}. The condition on $\nabla \cdot \vec{B}$ is controlled using the Powell 8-wave scheme \citep[][]{1999JCoPh.154..284P} employed in \cite{2013MNRAS.432..176P, 2018MNRAS.480.5113M}. In addition to ideal MHD, we include a CR component using the two-fluid
approximation implemented in AREPO \citep[][]{2016MNRAS.462.2603P,2017MNRAS.465.4500P}. The
CR component has an adiabatic index of $\gamma_{CR}=4/3$ and is injected as part of the jet. The complete set of equations of ideal MHD for a two-fluid medium composed of thermal gas and CRs can be found in Sec. 2.1 of \citealt{2017MNRAS.465.4500P}. The conservation laws of total energy density, $\varepsilon$, and CR energy density, $\varepsilon_{CR}$, used in this work are:
\begin{equation}
    \frac{\partial \varepsilon}{\partial t} + \vec{\nabla} \cdot \left( (\varepsilon + P) \vec{v} - \vec{B}(\vec{v} \cdot \vec{B})\right)
    = P_{CR} \vec{\nabla} \cdot \vec{v}, 
\end{equation}
\begin{equation}
    \frac{\partial \varepsilon_{CR}}{\partial t} + \vec{\nabla} \cdot \left( \varepsilon_{CR} \vec{v} \right)
    = - P_{CR} \vec{\nabla} \cdot \vec{v} + \Lambda_{CR},
\end{equation}
where $\Lambda_{CR}$ is a gain term for the CR energy density. The total pressure, $P$, and energy density, $\varepsilon$, are given by
\begin{equation}
    P=P_{th} + P_{CR} + \frac{\vec{B}^2}{2},
\end{equation}
\begin{equation}
    \varepsilon=\varepsilon_{th} + \frac{\rho \vec{v}^2}{2} + \frac{\vec{B}^2}{2}.
\end{equation}
Note in this work we closely follow the numerical setup in \citealt{2021ApJ...914...73Z}. We do not include the CR spatial diffusion and streaming terms. Hadronic interactions of CRs with thermal gas protons, Coulomb, and ionization interactions which result in CR energy losses are ignored in this work as well. The present work represents the first step in developing a comprehensive parameter space for central AGN and cluster merger characteristics. In a following paper, we will include more realistic models for CR dynamics.

The simulation domain is chosen to be a cubic box of size $L = 40$ Mpc with periodic boundary conditions. Nevertheless, in the remainder of the paper we will only present results of the inner $~10$ Mpc region. Following \cite{2021ApJ...914...73Z}, we ensure that the initial condition is free of spurious gas density and pressure fluctuations by performing a mesh relaxation step for $\sim 100$ timesteps before each simulation.
Each low-resolution (high-resolution) simulation initially has $10^6$ ($10^7$) gas cells and $10^6$ ($10^7$) DM particles. 
The gas cells are initialized with the same mass but are allowed to undergo mesh refinement and derefinement during the simulation.
The reference mass for our low-resolution (high-resolution) runs is $8.626 \times 10^7 \mathrm{M}_{\odot}$ ($8.626 \times 10^6 \mathrm{M}_{\odot}$). We perform a total of 48 simulations at $8.626 \times 10^7 \mathrm{M}_{\odot}$ resolution and we indicate which of these were also performed at $8.626 \times 10^6 \mathrm{M}_{\odot}$ resolution in Table~\ref{table:sims}.

\begin{figure}
    \centering
    \includegraphics[width=0.75\linewidth]{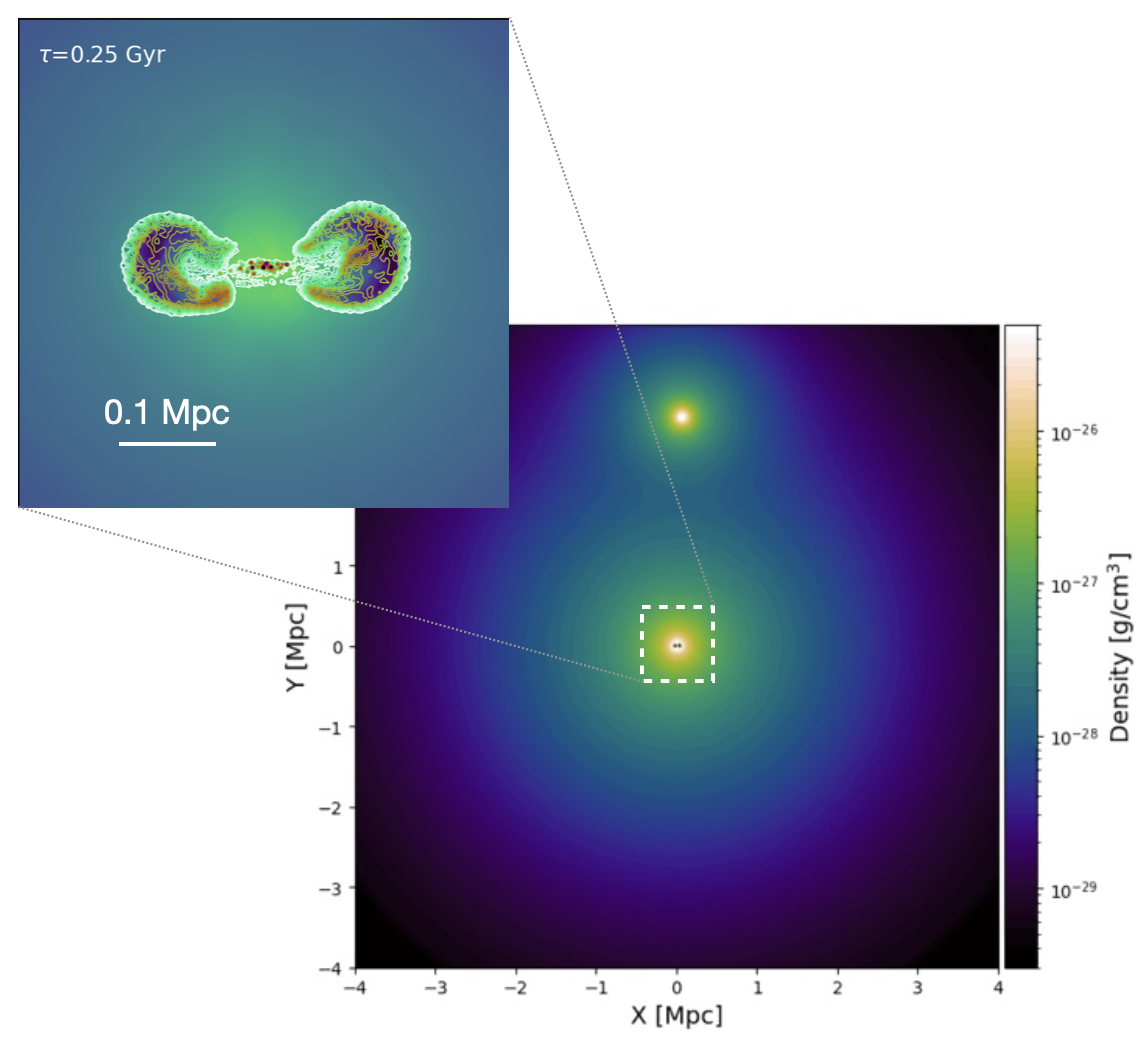} \\
    \includegraphics[width=0.75\linewidth]{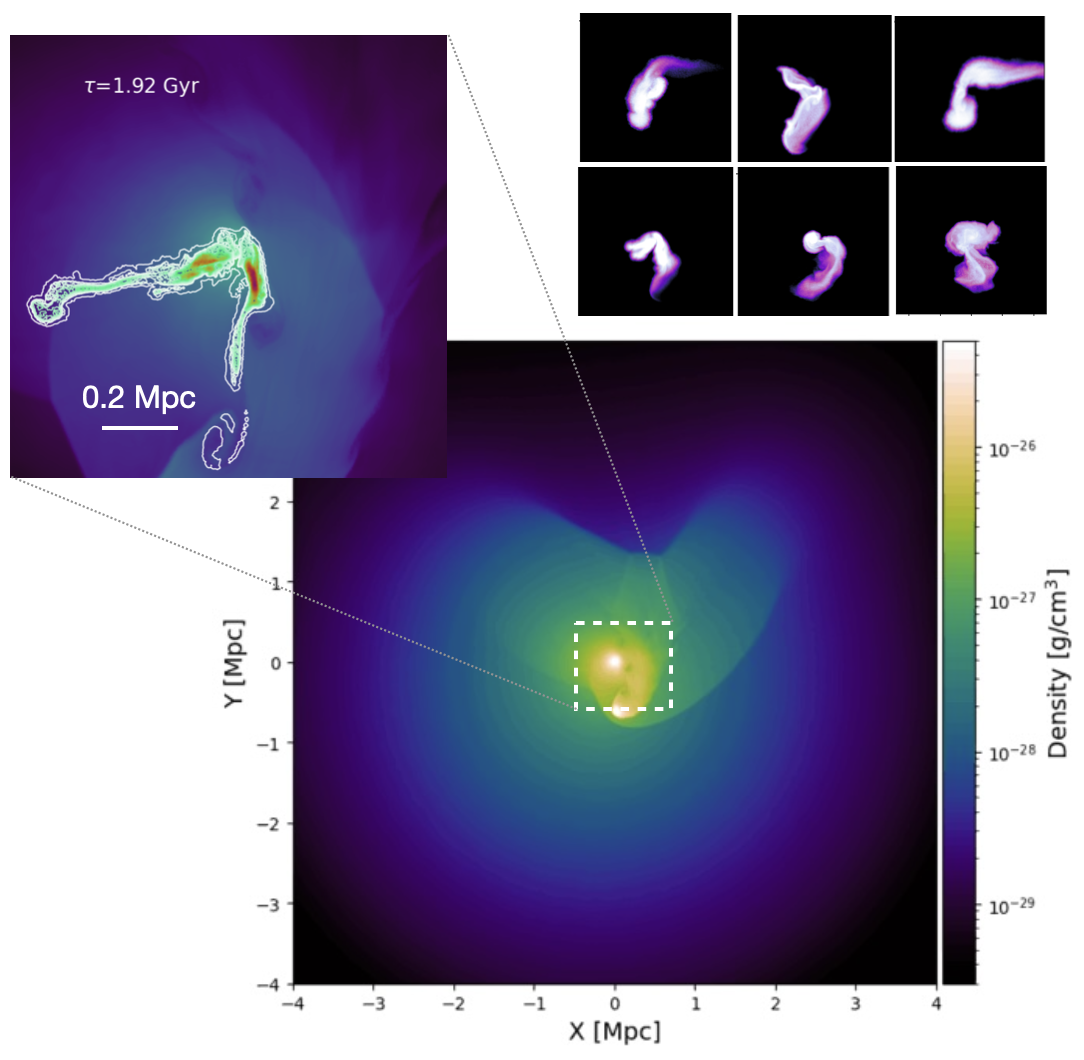} \\
    \includegraphics[width=0.75\linewidth]{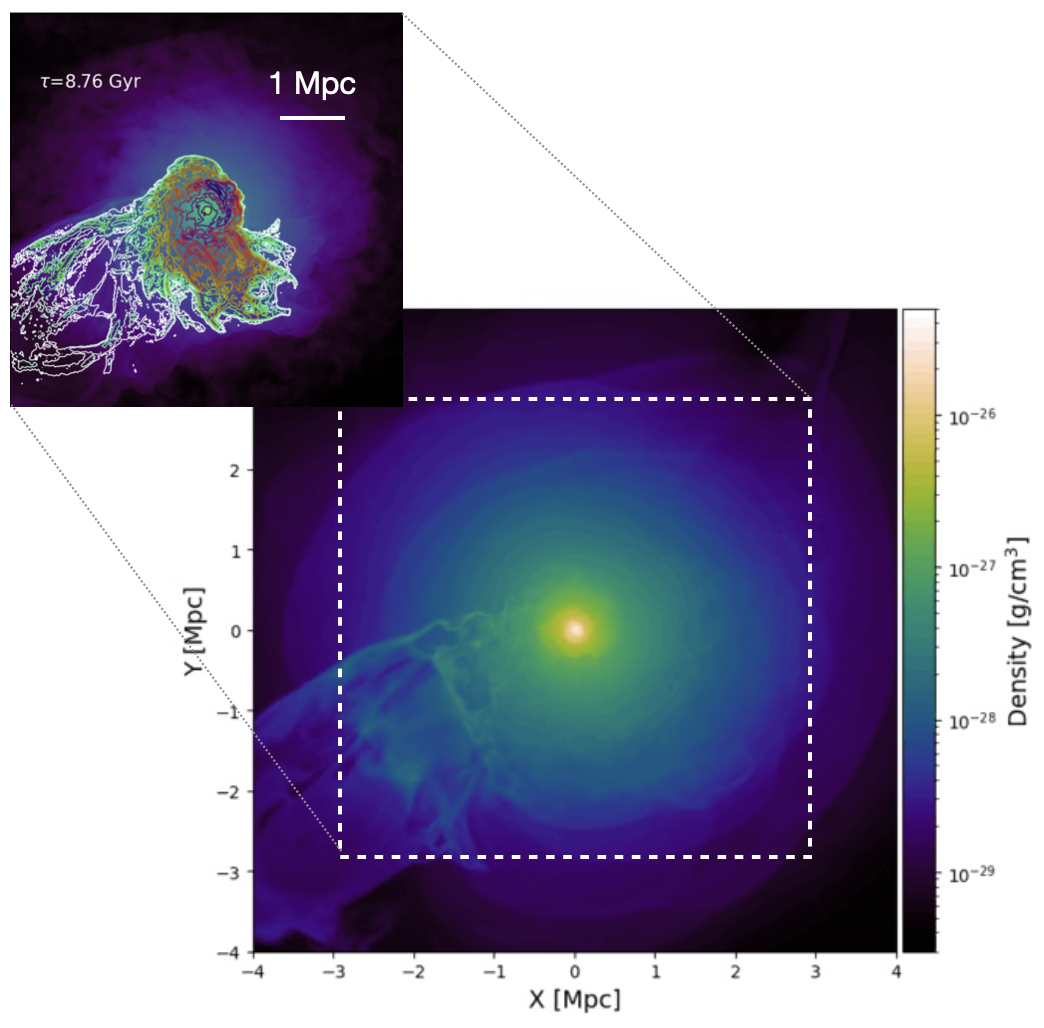}
    \caption{Visualization of the merger and AGN jet material evolution. The upper, middle and lower panels depict the early-, mid-, and late phases of run $R=1:5$, $\theta=20^{\circ}$, $t_{jet,0}=50$ Myr at high resolution. The lower panels show density projection maps with a width of 8 Mpc. The upper left insets show a zoom-in in selected areas (white dashed lines) where the AGN jet material is (contour lines). At the top of the middle panel, we additionally show six examples of different morphologies drawn by the CR jet pressure for different runs. The complete set of projection maps of the CR pressure with colorbars can be found in Appendix~\ref{app:theta10}.}
    \label{fig:new}
\end{figure}

We follow a similar jet simulation set-up to \citet{2021ApJ...914...73Z}. We provide a short summary in the following. We place a black hole particle at the center of the main galaxy cluster that has a mass of $\mathrm{M}_{\mathrm{BH}}=6.7 \times 10^8 \mathrm{M}_{\odot}$\footnote{The value
of the mass is of little importance because we do not perform accretion onto the
black hole particle.} and use the approach of \citet{2017MNRAS.470.4530W, Weinberger2023}. This method injects a bi-directional jet which is kinetically dominated, low density, and collimated. Kinetic, thermal, magnetic and CR energy are injected into two small
spherical regions a few kpc from the location of a black
hole particle. In our set-up, the black hole particle interacts gravitationally with the cluster's gravitational potential and it is not re-positioned at the potential's center\footnote{We checked that this does not affect the computation of the CR material radius (see Section~\ref{sec:evolution_jet_material}).}. We do not consider accretion onto the
black hole particle in our simulations. 
In each simulation, we considered a jet power of $P_{\mathrm{jet}} = 3 \times 10^{45} \, \mathrm{erg} \, \mathrm{s}^{-1}$ and a jet density of $\rho_{\mathrm{jet}}=1.51 \times 10^{-28} \, \mathrm{g} \, \mathrm{cm}^{-3}$. In all simulations, the jet is active for 100 Myr which corresponds to a total
injected energy of $E_{\mathrm{jet}} =9.46 \times 10^{60} \mathrm{erg}$. The runs are performed with a magnetized jet and the injected magnetic field is
purely toroidal. We considered equal magnetic and thermal pressures, $\beta_{\mathrm{jet}}=P_{th}/P_{B}=1$, and equal CR and thermal pressures, $\beta_{\mathrm{jet,CR}}=P_{th}/P_{CR}=1$, inside the jet regions. Most of the energy at the injection scale is added as kinetic energy and only thermalizes once the jet slows down, leading to highly thermally dominated lobes \citep{2017MNRAS.470.4530W}. To allow a fraction of $10\%$ of the injected jet energy to be deposited as cosmic rays in the lobes, we additionally inject a passive scalar that causes the jet material to gradually, over the characteristic time of $1.5$ Myr\footnote{Thus, the conversion from thermal energy to cosmic rays happens far more quickly than any of the relevant timescales of a galaxy cluster.}, convert the desired amount of thermal to cosmic ray energy, leading to substantial fractions of cosmic rays in the lobes. The precise algorithm will be detailed in a forthcoming paper (Weinberger et al., in prep.). We explored different orientations of the jet axis and different times for the jet initialization, $t_{\mathrm{jet},0}=50$ Myr, 1 Gyr, 2 Gyr, 5 Gyr. Finally, all our simulations are run for 9 Gyr.

In Table~\ref{table:sims} we summarize the simulations considered for this work.

\newpage

\section{Results}
\label{sec:results}

\subsection{Merger and jet characteristics}
\label{sec:merger_charac}

%
In Fig.~\ref{fig:new}, we summarize the main phases of the AGN jet material during the evolution of the merger for a representative run. We refer the reader to Appendix~\ref{app:theta10} where we show projection plots of various simulations listed in Table~\ref{table:sims}. In each figure in Appendix~\ref{app:theta10}, the reader will find density and CR pressure projection maps.

The binary merger evolves as follows: initially, the infall of the secondary cluster onto the primary sets up gas sloshing of the central core within the dark-matter-dominated potential. 
After the first core passage, the secondary cluster starts to lose mass and momentum as it is stripped of gas and DM by ram-pressure and dynamical friction.
The core of the secondary then drifts to the outskirts of the primary cluster (approximately 1--2 Mpc away from the main core, depending on the mass ratio) before gravitating back towards the center. During this journey, gas stripped from the secondary blends with the primary cluster's gas through mixing driven by fluid instabilities and turbulence.
The final phase occurs after the second core passage and it culminates in the complete disruption of the secondary's core. This process is very similar for both mass ratios \citep[see][for previous studies for off-center, unequal-mass cases]{2001ApJ...561..621R,2006MNRAS.373..881P,2011ApJ...728...54Z,2019ApJ...883..118B}.

\begin{figure}
    \includegraphics[width=0.5\textwidth]{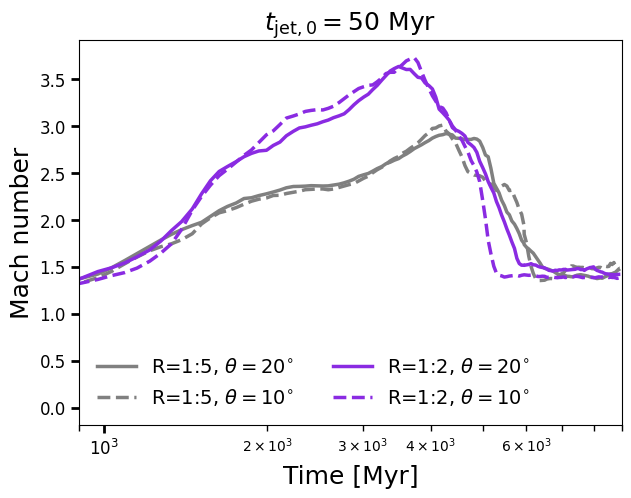}
    \includegraphics[width=0.5\textwidth]{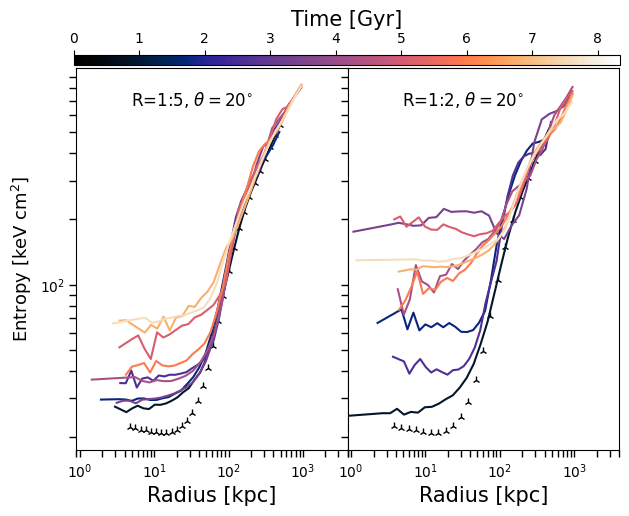}
    \caption{Upper panel: Evolution of the volume-weighted mean Mach number inside the central region ($r<5$ Mpc) of the main cluster for both mass ratios and impact parameters. Lower panel: Mass-weighted entropy radial profiles as a function of time. The profiles are computed with a sphere with radius $1$ Mpc. The markers indicate the initial profile of the central cluster. We only show two representative cases where the jet direction is initialized in the $x$-direction.}
    \label{fig:pdf_mach}
\end{figure}

The evolution of the binary merger significantly influences the dispersal of CRs injected by the jets. We identify three distinct phases.  In the \textit{early phase} ($\tau \lesssim 0.5$ Gyr\footnote{Note that $\tau=t - t_{\mathrm{jet},0}$.}), the jets form cavities within the ICM density. 
For both examined mass ratios, these cavities appear less disturbed at $t_{\textrm{jet},0}=50$ Myr and $t_{\textrm{jet},0}=1$ Gyr due to the secondary cluster not yet completing its initial core passage. Conversely, cavities produced by jet bursts occurring post-first core passage tend to be disrupted more noticeably and quickly. In the $R=1:5$ cases, none or only one cavity is distinctly visible (see Fig.~\ref{fig:maps_r0p2} in Appendix~\ref{app:theta10}). While this will be subject to future work, it is worth noting that the environmental disruption of the AGN lobes is likely also common in galaxy groups \citep[see][for observational examples]{2011ApJ...732...95G,2022A&A...661A..92B}. 
During the \textit{mid-phase} ($1.5 \lesssim \tau \lesssim 3$ Gyr), the jet lobes expand, creating diverse morphologies, including wide-angle tailed (WAT)-like or head tailed-like structures (see Fig.~\ref{fig:new}), middle panel, or Figs.~\ref{fig:maps_r0p2} and \ref{fig:maps_r0p5_theta10} in Appendix~\ref{app:theta10}. 
We will further discuss the morphology of specific cases and snapshots in Section~\ref{sec:sources}.  Finally, in the \textit{later phase} ($\tau \gtrsim 5.5$ Gyr), the jet remnant predominantly occupies the main cluster's center with a rounded shape. By the end of the simulation, the jet material covers a projected linear size extending up to approximately 2-4 Mpc. Notably, the $R=1:5$ case with $t_{\textrm{jet},0}=50$ Myr reaches a projected linear extent of about 6 Mpc (see Fig.~\ref{fig:new}, lower panel). 
The binary merger process naturally generates shocks within the ICM of the main cluster. Using the AREPO shock-finder \citep[][]{2015MNRAS.446.3992S}, we identify the shocked regions. 
The upper panel of Fig.~\ref{fig:pdf_mach} shows the evolution of the volume-weighted mean Mach number for our two mass-ratio and two impact- parameter cases. We considered only those shocked cells with $\mathcal{M}>0$. The merger with a mass ratio of $R=1:2$ induces slightly higher Mach numbers compared to the $R=1:5$ merger. This is true for both $\theta=20^{\circ}$ and $\theta=10^{\circ}$, and is mainly due to the fact that the final mass is larger in the systems with $R =1:2$.
The first shocks typically begin to form when the core of the secondary cluster is within $\sim$1.5 Mpc of the main cluster's core. Each core passage generates shock waves that propagate from the inner region towards the outskirts. Towards the end of the simulation, as the cluster settles into a more relaxed state, the strength of the shocks diminishes, showing Mach numbers in the range of $\mathcal{M}\sim 1$--2. 
A comparison of the mean Mach number for different initial jet orientations was not presented in Fig.~\ref{fig:pdf_mach}, as the differences with respect to the runs with the jet initialized in the $x$-direction are not significant. This implies that the shocks in our simulations primarily originate from the merger rather than the AGN activity.
In the following section and Section~\ref{sec:relics}, we will focus more on the interaction between these shocks and the CR plasma ejected by the AGN.

In the lower panel of Fig.~\ref{fig:pdf_mach}, we show the evolution of the entropy profiles within a 1 Mpc radius.  
We show two cases with different mass ratios and $\theta=20^{\circ}$ in different panels. We note that the cases with $\theta=10^{\circ}$ produce similar entropy profiles and are therefore not shown in Fig.~\ref{fig:pdf_mach}.
In the later stages of the simulation, $t\gtrsim 4$ Gyr, there is a marked increase in entropy at $r\lesssim 200$ kpc, which is influenced by the initial merger conditions. We suspect that the level of this increase that is notably dependent on the mass ratio of the merger is due to mixing \citep[see e.g. discussions in][for similar results from adiabatic simulations]{2009MNRAS.395..180M,2011ApJ...728...54Z,2021MNRAS.504.5409V}.

\begin{figure*}
    \centering

    \includegraphics[width=0.85\textwidth]{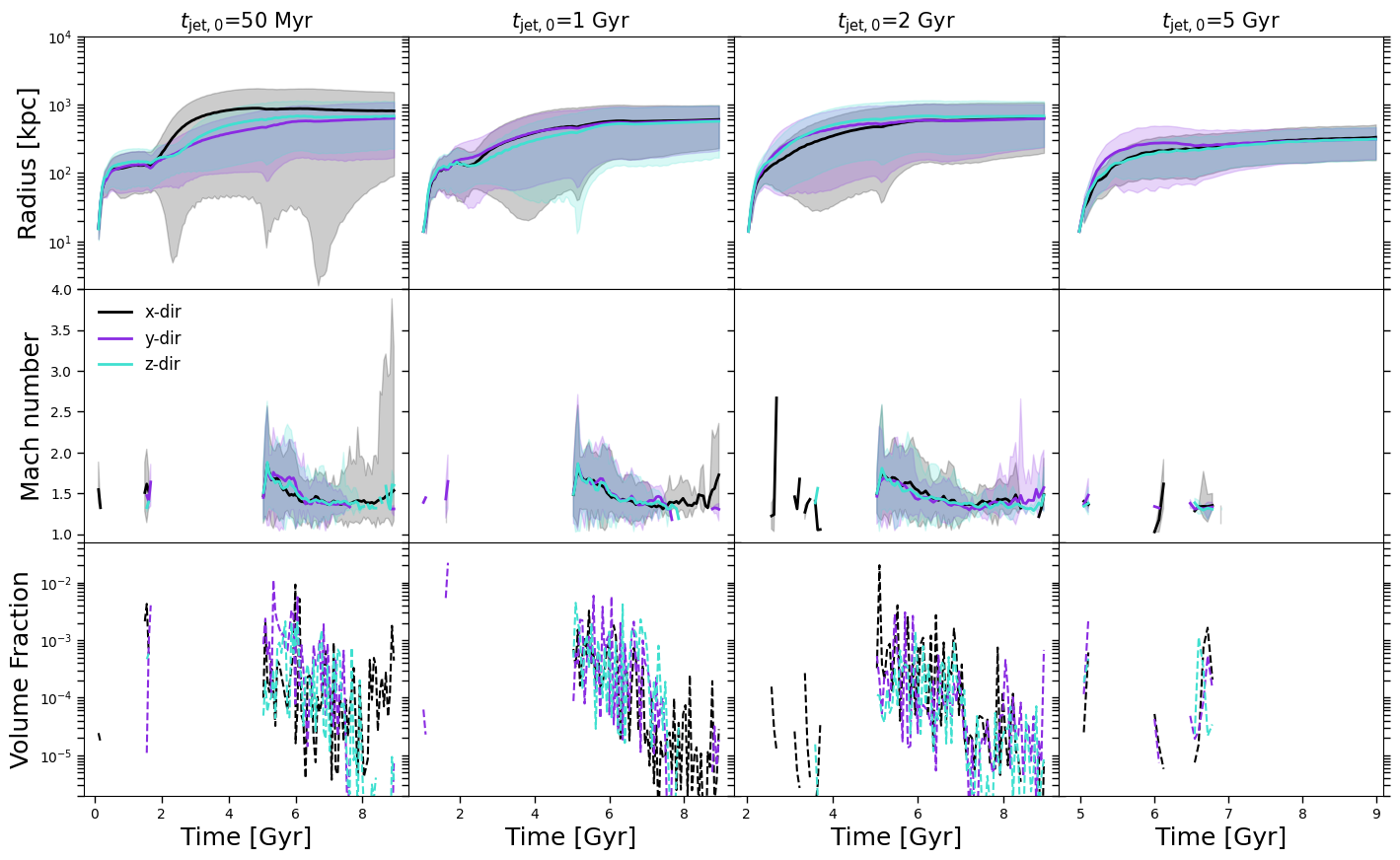}
    \includegraphics[width=0.85\textwidth]{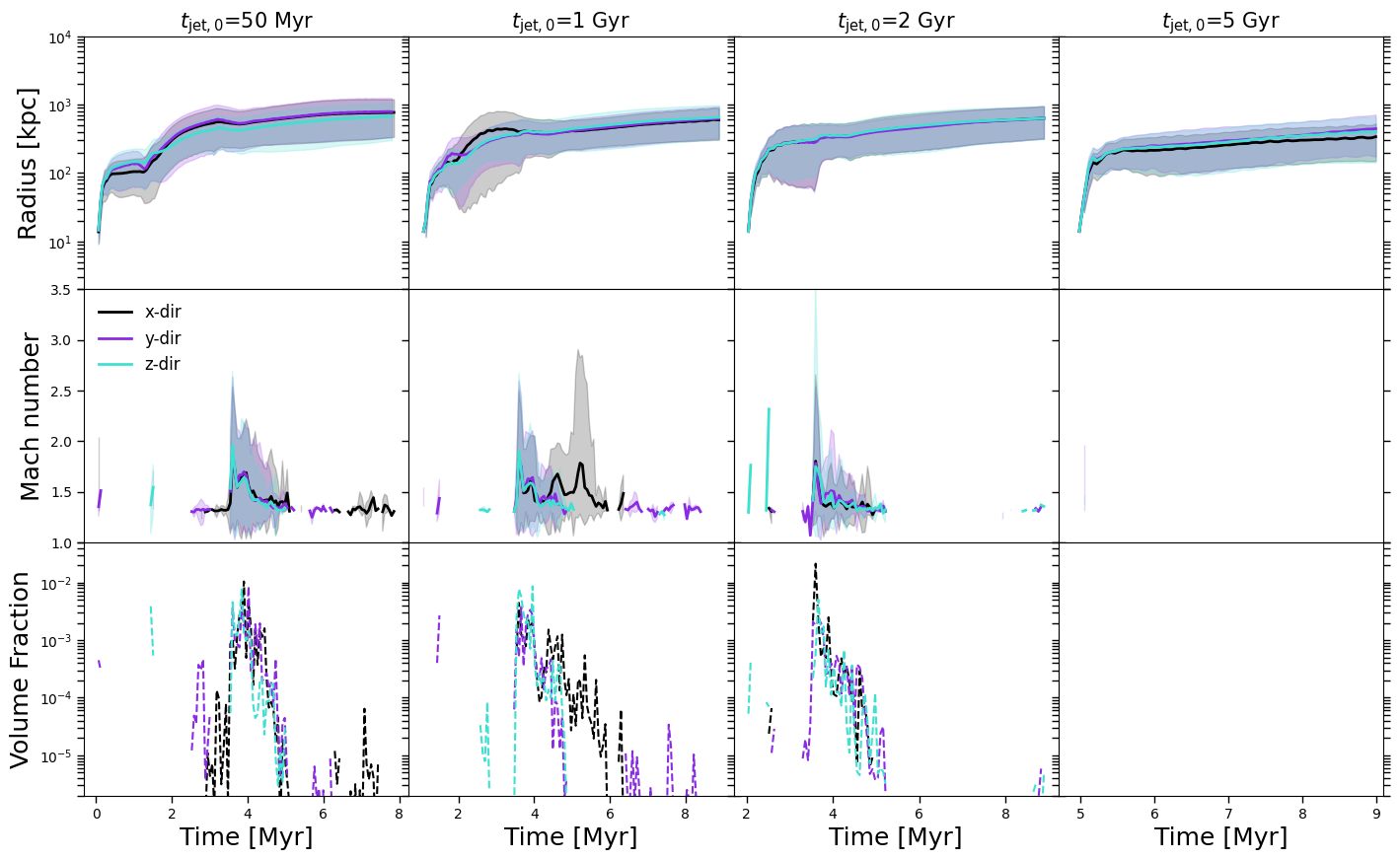}
    \caption{ First row: Volume-averaged radial extent of jet material centered on the black hole particle coordinates (solid lines), with shaded areas representing $\pm 1$ standard deviation. Second row: Volume-averaged Mach number of the shocked cells within the jet material, with shaded areas indicating the range from minimum to maximum values. Third row: Volume fraction of the shocked jet material relative to the total jet material volume. The figure's upper and lower panel sets correspond to the mass ratios R=1:5 and R=1:2, respectively, both with an impact parameter of $\theta=20^{\circ}$. 
    }
    \label{fig:multipanel_evolution}
\end{figure*}

\begin{figure*}
    \centering

    \includegraphics[width=0.85\textwidth]{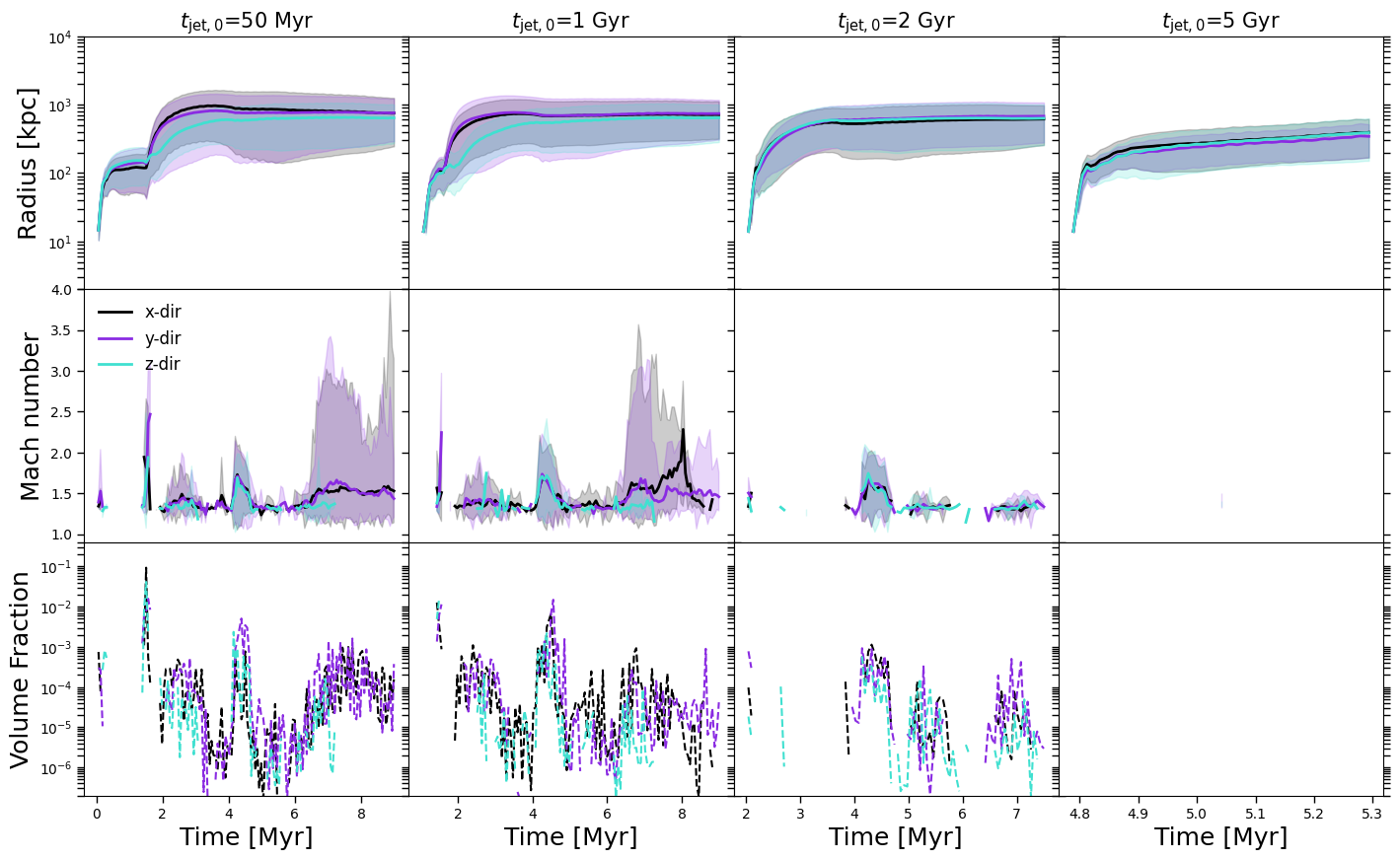}
    \includegraphics[width=0.85\textwidth]{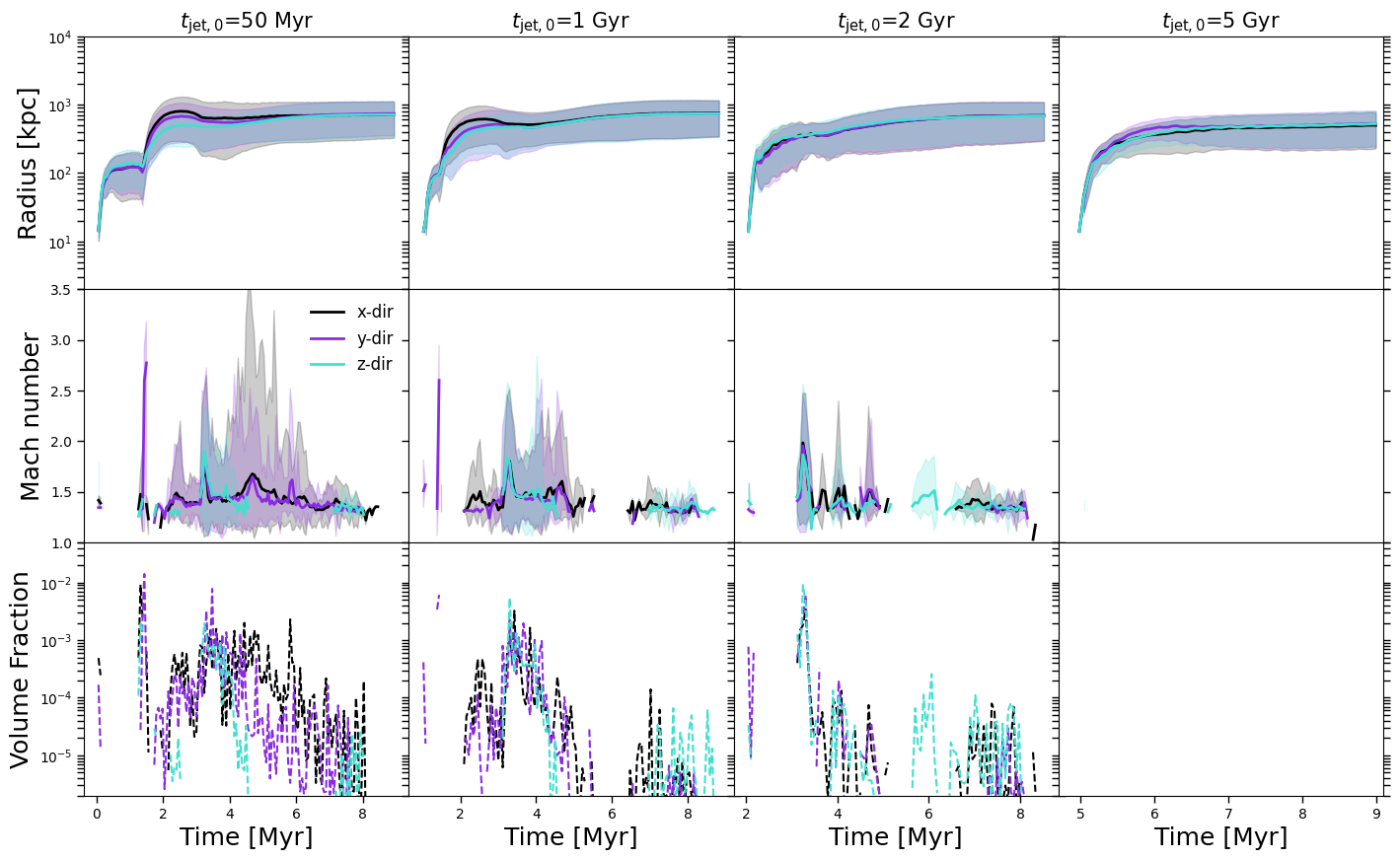}
    \caption{Same as Fig.~\ref{fig:multipanel_evolution} but for the runs with $\theta=10^{\circ}$.
    }
    \label{fig:multipanel_evolution_theta10}
\end{figure*}

\subsection{Evolution of jet material}
\label{sec:evolution_jet_material}

In this section, we describe more in detail the evolution of the jet material throughout the merger event. Fig.~\ref{fig:multipanel_evolution} and Fig.~\ref{fig:multipanel_evolution_theta10} illustrate the temporal progression of this material\footnote{For a detailed explanation of how we tracked the jet material, refer to Appendix~\ref{app:jet_material}.} in all runs with $\theta=20^{\circ}$ and $\theta=10^{\circ}$, respectively. 
In the first rows, we show the volume-averaged radius reached by the jet material. We considered the coordinates of the black-hole particle as the center for computing this radial extent. As discussed in Section~\ref{sec:merger_charac} (see Fig.~\ref{fig:pdf_mach}), we keep track of the shocked regions/cells with the AREPO shock-finder \citep[][]{2015MNRAS.446.3992S} throughout the simulations. These shocks induced by the binary merger, also propagate through the jet material. In the second rows of Fig.~\ref{fig:multipanel_evolution} and Fig.~\ref{fig:multipanel_evolution_theta10} we show the mean Mach number corresponding to those shocked cells that overlap with the CR jet material. In the third rows, we show the volume fraction of the shocked jet material relative to the total jet material volume.

We see that the CR material can reach very large radii, up to 1-2 Mpc. Rapid increments occur at the times of the first and second core passages. This is especially the case for cases with $t_{\mathrm{jet,0}}=50$ Myr, where the maximum radial extend of the CR jet material is observed at $t\sim 2$ Gyr, corresponding to the time of the second core passage. This is also the case for runs with $\theta=10^{\circ}$ shown in Fig.~\ref{fig:multipanel_evolution_theta10}, suggesting that the final extension of the CR material is independent of the initial $\theta$ or impact parameter. Similarly, the jet's initial direction—whether in the $x$-, $y$-, or $z$-axis—does not significantly affect the final mean radius of the distribution. On the other hand, we find that the final mean radii depend on the launching time of the jets, $t_{\mathrm{jet,0}}$ (compare columns in Figs.~\ref{fig:multipanel_evolution}--\ref{fig:multipanel_evolution_theta10}). The earlier the jet outburst occurs, the larger the final radial extent. For example, in the cases where $t_{\mathrm{jet,0}}=5$ Gyr, the radial extent reaches only $\sim 300$--500 kpc. This suggests that the combination of the mass ratio, and thus the merger history, along with the launching time of the jets primarily influence the maximum spread of CRs coming from AGN bubbles.

Note that during the initial disruption of the lobes, these three parameters, $\theta$, $t_{\mathrm{jet,0}}$ and the initial direction of the jets, significantly impact the early disruption and resultant morphology of the lobes, as we will further explore in Section~\ref{sec:sources}.

We note that, contrary to the shock statistics in the whole central volume ($r<5$ Mpc, see Fig.~\ref{fig:pdf_mach}), the mean Mach numbers detected only in the jet-material regions (see second rows of Figs.~\ref{fig:multipanel_evolution}--\ref{fig:multipanel_evolution_theta10})
are typically low and depending on $\theta$ and R. In the $\theta=20^{\circ}$ runs, the mean Mach number is especially low, $\sim$1.5--2, in both mass ratio cases. Spurious large Mach numbers are detected for example in the $t_{\mathrm{jet,0}}=50$ Myr, $x$-direction, $R=1:5$ run, reaching $\mathcal{M}\sim 4$, and the $t_{\mathrm{jet,0}}=2$ Gyr, $z$-direction, $R=1:2$ run reaching $\mathcal{M}\sim 3.5$. This picture is slightly different for the $\theta=10^{\circ}$ runs where the mean shock Mach number ranges between $\mathcal{M}\sim 1.5$ and $\mathcal{M}\sim 2.7$ for both mass ratios. We note that a burst episode taking place at the late binary merger evolution is less prone to interact with any merger shock. In this case, a late time would correspond to a time past the second core passage. This can be seen in the fourth columns of Figs.~\ref{fig:multipanel_evolution}--\ref{fig:multipanel_evolution_theta10} where we show the results from the $t_{\mathrm{jet,0}}=5$ Gyr runs.
Additionally, a smaller initial impact parameter, such as $\theta=10^\circ$, results in more extended periods of interaction between the jet material and merger shock.
For instance, while the $\theta=20^\circ$, $t_{\mathrm{jet,0}}=50$ Myr runs exhibit a jet-shock interaction period lasting for about 4 Gyr, the $\theta=10^\circ$, $t_{\mathrm{jet,0}}=50$ Myr runs exhibit more prolonged periods that can extend to $\sim$7 Gyr. This is of particular relevance for the possible formation of radio relics through the re-acceleration of fossil electrons via DSA. These results indicate that smaller initial impact parameters increase the likelihood of the jet material being shocked by the merger, thereby enhancing the potential for forming radio relics. 

Finally, in all our runs, the fraction of CR material that encounters a merger shock amounts to $\lesssim 1$\%. The only exception is the R=1:5, $\theta=10^\circ$, $t_{\mathrm{jet,0}}=50$ Myr run, where $\sim 10$\% of the CR material is shocked at $t\sim 1.5$ Myr.

We refer the reader to Appendix~\ref{app:high-res} for a comparison of these evolutionary trends at higher resolution (see simulations that were performed at $8.626 \times 10^6~\mathrm{M}_{\odot}$ in Table~\ref{table:sims}). In that Appendix we mainly discuss the results of the $R=1:5$, $\theta=20^{\circ}$ runs where we set the jet to ignite in the $x$-direction and compare to the results shown in the upper panels of Fig.~\ref{fig:multipanel_evolution}. We show that the evolutionary trends discussed in this section are consistent at higher resolution.

%

\subsection{Mixing of the AGN bubble with the ICM}

\begin{figure}
    \centering

    \includegraphics[width=0.8\columnwidth]{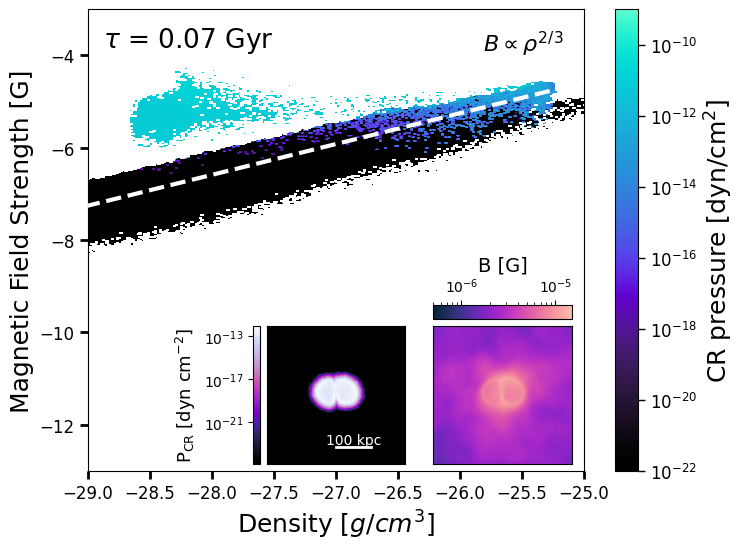}
    \includegraphics[width=0.8\columnwidth]{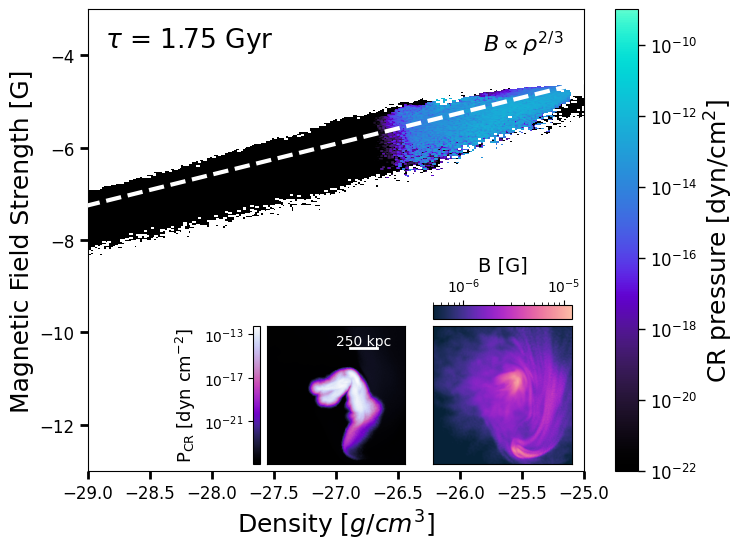}
\includegraphics[width=0.8\columnwidth]{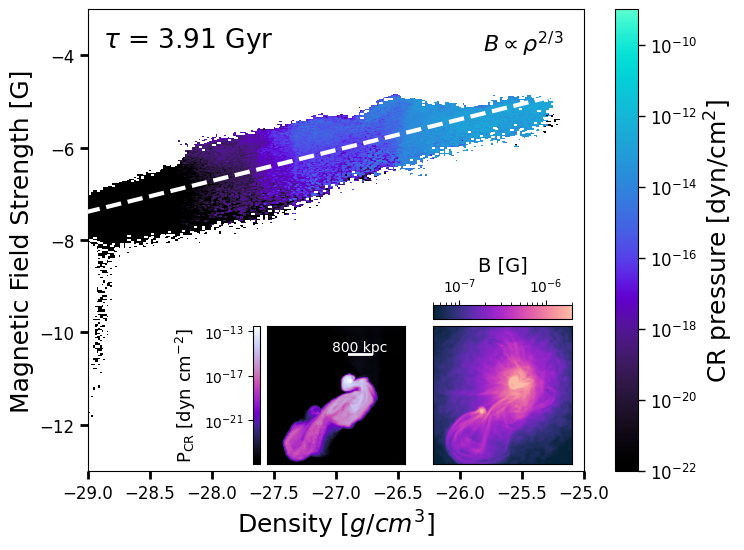}
    \includegraphics[width=0.8\columnwidth]{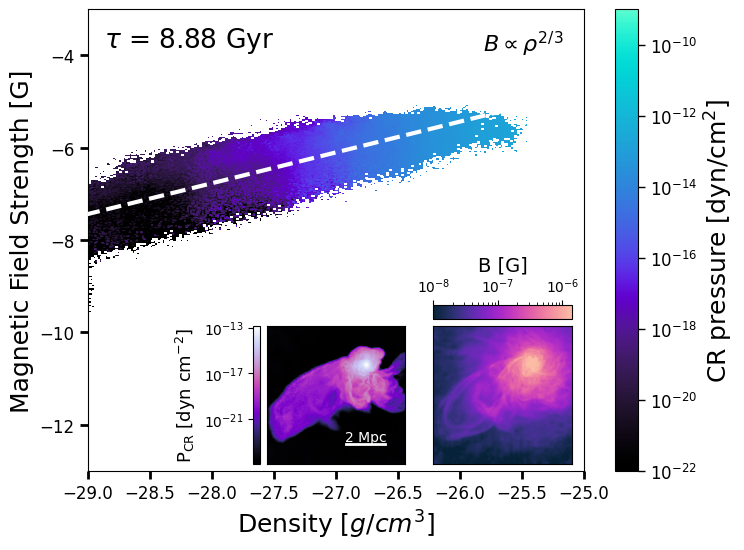}
    \caption{2D Histogram of the magnetic field and density weighted by the CR pressure. The insets correspond to projection maps of the CR pressure (left) and magnetic field strength (right). The dotted white line shows the the expected scaling based on
adiabatic compression. }
    \label{fig:phase-plots}
\end{figure}

In Fig.~\ref{fig:phase-plots}, we present phase plots of the magnetic field and density from the R2\_jetdir-x\_50Myr run. The left-hand side illustrates values derived from 3D data. Data points are color-coded based on CR pressure. Through these plots, we can trace the dynamics of jet plasma mixing with the ICM. 
In the initial stage, CRs remain tightly contained within the bubbles, showing no signs of mixing with the ICM, indicated by the clear separation between cyan and black points. The subsequent phase begins approximately 1.7 Gyr after the jet is ignited, coinciding with the first core passage. This marks the onset of bubble disruption and the commencement of material mixing with the ICM. This mixing process extends until $\sim$4 Gyr post-jet ignition, characterized by progressively lower CR pressure values in regions of lower density that are also weakly magnetized. It is important to note that during these mixing intervals, regions with a stronger magnetic field do not always correspond to higher CR energy, as illustrated by the purple points in the second and third rows of Fig.~\ref{fig:phase-plots}. 
In the final stage, occurring $\sim$9 Gyr after the jet ignition, the material has completely integrated with the ICM. This is evident from Fig.~\ref{fig:new}, lower panel and all figures in Appendix~\ref{app:theta10}, 
where the material adopts a rounded shape, occupying the central $\sim$1--2 Mpc of the cluster. During this last phase, a more pronounced correlation emerges between regions of stronger magnetic fields and higher CR pressure.

\section{Discussion}

\subsection{Resemblance with observed radio sources}
\label{sec:sources}

In this section, we briefly explore the formation of various radio sources and their connection to the initial conditions of the merger and the jet ignition. We concentrate on the early development phase of the jet material, prior to its widespread dispersal throughout the central region of the main cluster, as detailed in Section~\ref{sec:evolution_jet_material}.

\begin{figure}
    \centering
    \includegraphics[width=0.8\linewidth]{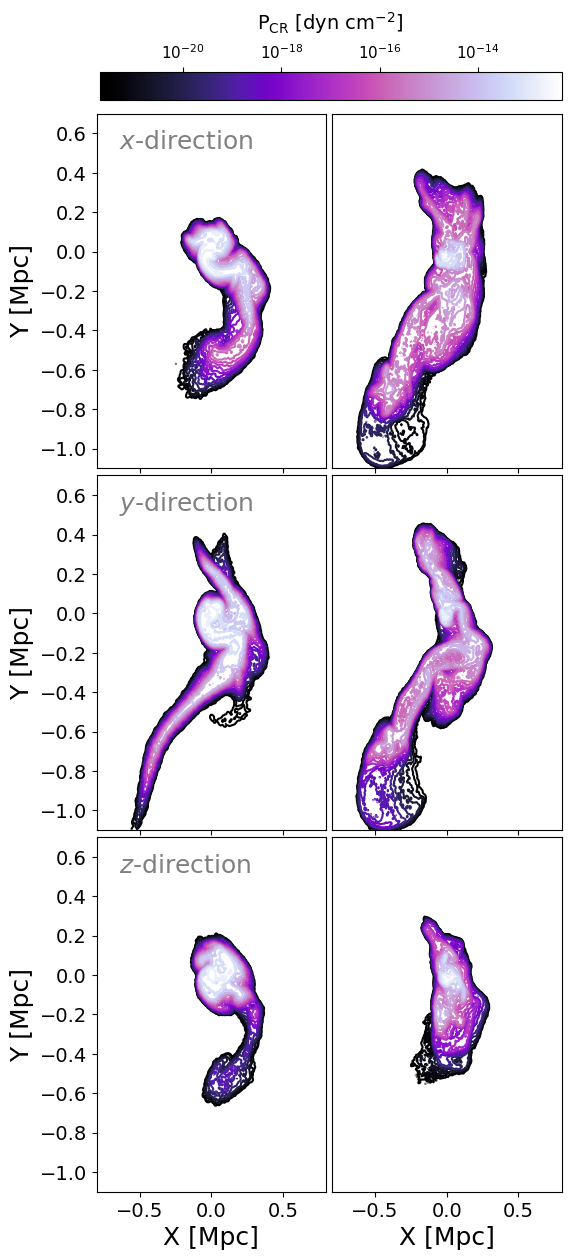}
    \caption{CR pressure contour maps of runs $R=1:5$, $t_{jet,0}=50$ Myr with $\theta=20^{\circ}$ (left column) and $\theta=10^{\circ}$ (right column) at $\tau=1.77$ Gyr. Each row shows morphological differences depending on the initial direction of the jet.}
    \label{fig:jetdir}
\end{figure}

\begin{itemize}
    \item[i)]\textit{Wide-Angle-Tail (WAT) sources}: 
    Our simulations indicate that the formation of sources like those observed is largely influenced by bulk motions. Specifically, we observe the emergence of WAT-like sources \citep[see][and references therein]{2023Galax..11...67O} with tails extending tens of kpc, starting around $\tau \sim 1.5$ Gyr, which correlates closely with the timing of the first core passage 
    (see Fig.~\ref{fig:new}, middle panel, and second columns of Figs.~\ref{fig:maps_r0p5}, ~\ref{fig:maps_r0p5_theta10} and ~\ref{fig:maps_r0p2_theta10} in Appendix~\ref{app:theta10}). These sources arise regardless of the merger's mass ratio but depend on the jet’s initial burst timing, the initial direction of the jet (see Fig.~\ref{fig:jetdir}),  and the position of the minor cluster, only forming if the burst precedes the first core passage. WATs have been found in both cool- and non-cool-core clusters \citep[see][]{2012PhDT.......292D,2015IAUS..313..315B}. Examples such as Abell 2029 \citep[][]{2013ApJ...773..114P} and Abell 1763 \citep[][]{2018ApJ...868..121D} showcase WATs in sloshing cool-core clusters, which would be more relevant for comparison with our current simulations. As shown in Fig.~\ref{fig:pdf_mach}, our simulations initiate within a cool-core setup, with AGN activity affecting only the central $\lesssim 20$ kpc region for less than $2$ Gyr. After this period, the cluster maintains lower entropy levels at its center compared to the outskirts. 
      \item[ii)]\textit{X-shaped sources}: 
        In our simulations, X-shaped radio sources resembling the observed pair of weak secondary lobes that form a cross-like shape \citep{1984MNRAS.210..929L} are found under specific conditions. These configurations appear in cases with R=1:2 and $t_{\mathrm{jet,0}}= 2$ Gyr, particularly when $\theta=20^{\circ}$ and $\theta=10^{\circ}$ (see Fig.~\ref{fig:maps_r0p5} and Fig.~\ref{fig:maps_r0p5_theta10} in Appendix~\ref{app:theta10}). This indicates that the formation of such structures may require a high mass ratio cluster merger and jet activity post-first core passage. While our simulations do not precisely mimic classic X-shaped radio galaxies like PKS 2014-55 \citep{2020MNRAS.495.1271C}, they do resemble others such as the radio galaxy in Abell 3670 \citep{2019A&A...631A.173B}. 
        Several studies propose alternative mechanisms for producing these sources, including jet precession \citep{2018MNRAS.474L..81L,2023ApJ...948...25N}, axisymmetric systems with intermittent jets \citep{2022ApJ...936L...5L}, and jet backflows altered by the surrounding environment conditions \citep{2011ApJ...733...58H} among others \citep[for a comprehensive review, see][]{2024FrASS..1171101G}. Observationally, it has been pointed out that the environment related to a cluster or group merger could be of relevance for the observed radio morphology \citep[][]{2019MNRAS.488.3416H}. Further analysis is necessary to specifically target these phenomena, but our results indicate potential links between cluster mergers and the emergence of X-shaped radio sources.

    \item[iii)]\textit{Wobbling/disrupted tails}: 
    Disrupted tails from WAT and hybrid sources are occasionally seen in observations. For instance, the radio galaxy MRC 0600-399 in Abell 3376 \citep{2021Natur.593...47C} and source J142914.03+465611.32 from the LOFAR Two-metre Sky Survey (LoTSS) \citep{2017A&A...598A.104S, 2019A&A...622A...1S}, discussed in \citealt{2020A&A...635A...5D}, both display tails that are distinctly disrupted. Notably, the end of one tail in these observations bends nearly perpendicularly to the jet direction, exhibiting a wobbling morphology. This is similar to what we simulate in scenarios with R=1:5, $\theta=20^{\circ}$, and $t_{\mathrm{jet},0}= 5$ Gyr, as shown in
    Fig.~\ref{fig:maps_r0p2} in Appendix~\ref{app:theta10}.
    Our findings suggest that a minor merger passing above the central black hole $\sim 0.4$ Gyr after jet ignition can sufficiently alter the tail's trajectory, resulting in noticeable wobbling substructures.

\end{itemize}

In summary, during this initial phase, we observe a variety of morphologies drawn by the jet CR material. What ultimately defines these different morphologies is an interplay between the stage of the merger, the initial jet direction, and the timing of the outburst. We note that the initial jet direction and the timing of the outburst are only relevant in these early and mid-phases.
For example, it is evident from the second column in Figs. 1--4 and all figures in Appendix A that different initial jet directions do not necessarily lead to the same morphologies. In the later phase, as discussed in Sec.~\ref{sec:merger_charac}, these parameters are of less relevance for the final shape of the CR material.
A comprehensive analysis of the formation of these different sources exceeds the scope of this paper and will be addressed in future studies.
%

\subsection{Contribution to radio relic emission?}
\label{sec:relics}

\begin{figure*}
    \centering

    \includegraphics[width=0.44\textwidth]{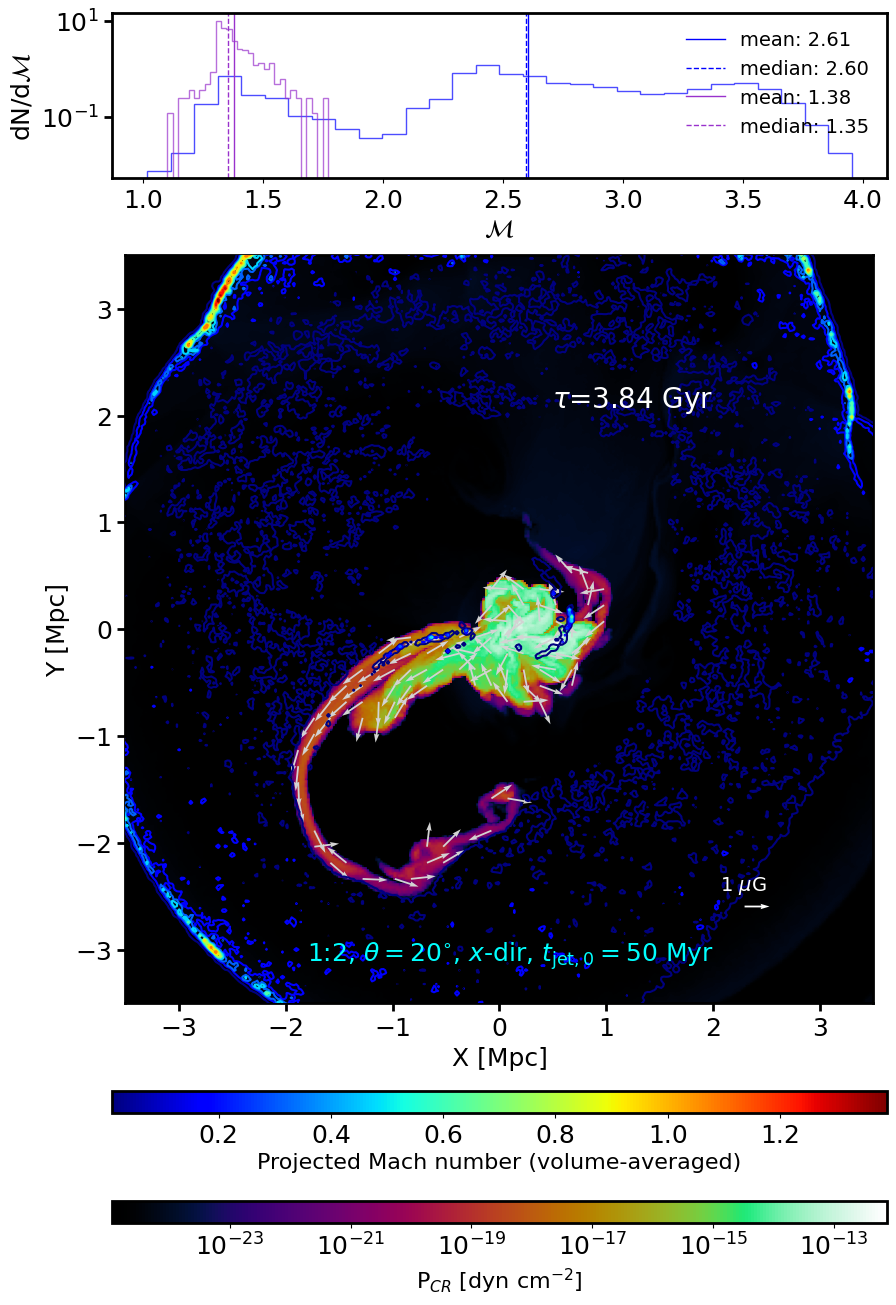}
    \includegraphics[width=0.48\textwidth]{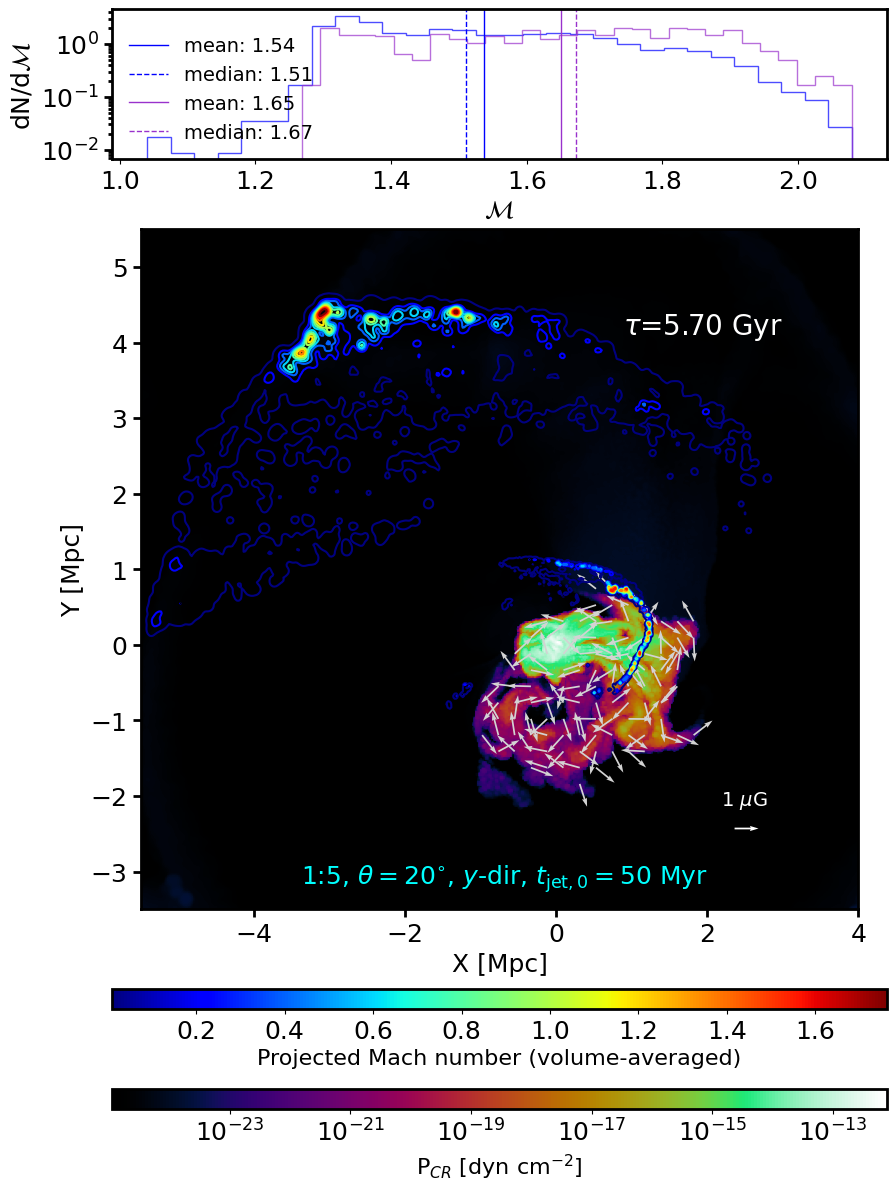}
    \includegraphics[width=0.475\textwidth]{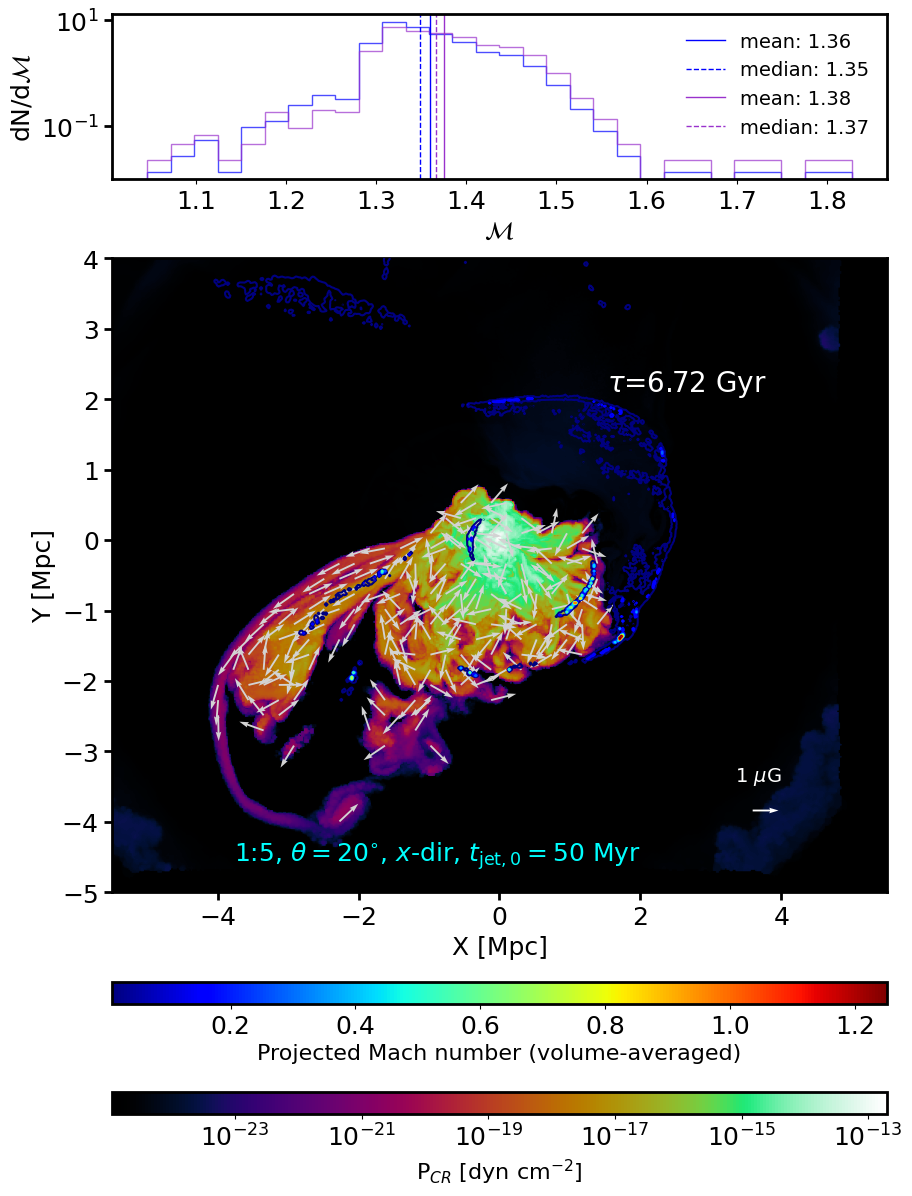}
    \includegraphics[width=0.47\textwidth]{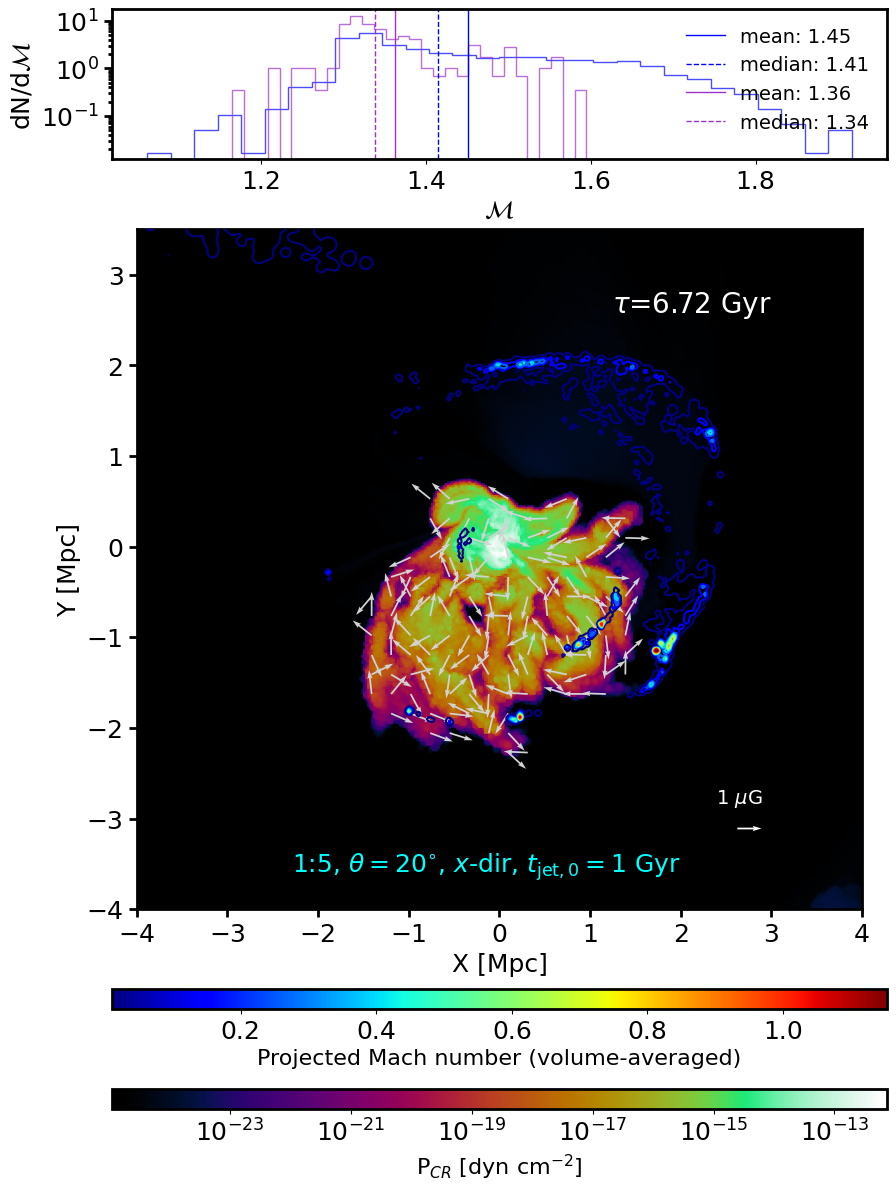}
    \caption{Thin CR pressure slices through the center of the main cluster. We overplot the Mach number contours. The upper panels show in blue the Mach number distribution of the whole shock that is crossing the CR jet material and in purple the Mach number distribution corresponding to the region that overlaps with the CR material. The magnetic field vectors are weighted with $B^2 \rho_{CR}$.}
    \label{fig:relics}
\end{figure*}

In this section, we show some examples of merger shocks interacting with the CR material (see Fig.~\ref{fig:relics}) and we discuss its potential to contribute to radio relic emission. As previously discussed in Section~\ref{sec:merger_charac}, during the merger, each core passage generates shocks with Mach numbers typically ranging from 2 to 4. Specifically, during the period from $t\sim2$--5 Gyr, the peak of the Mach number distribution is $\sim\mathcal{M}\sim2$--3 for a merger mass ratio of R=1:5 and $\mathcal{M}\sim3$--4 for a mass ratio of R=1:2 (see Fig.~\ref{fig:pdf_mach}).
 Later on, as the sub-cluster loses mass and momentum, the shocks in the main cluster ICM are weaker and thus their Mach numbers are lower.
 In Fig.~\ref{fig:relics}, we show one example of the CR jet material interacting with shocks at an early phase (see upper left panel in Fig~\ref{fig:relics}), and three other examples where the now more dispersed CR jet material interacts with the weaker shocks generated during the last phase of the merger. We overplot Mach number contours to visualize the location of the shocks.

 The underlying cluster turbulence naturally induces substructure along the shock fronts. We expect that this substructure will be also present in the synchrotron emission \citep[e.g.,][]{2019ApJ...883..138R,dominguezfernandez2020morphology,dominguez2021,Wittor_2021}. Yet, the observed radio surface brightness variations could differ in morphology depending on whether the thermal or fossil electron population participates in the DSA mechanism \citep[see][]{dominguez2024patchiness}. In Fig.~\ref{fig:relics}, the thin slices show the alignment of the magnetic field vectors predominantly along the bulk motions that impact and shape the CR jet material. Nevertheless, these vectors do not clearly align along the shock fronts, primarily due to the low Mach numbers of these shocks. Specifically, in MHD shocks, the magnetic field component parallel to the shock normal is conserved across the shock. However, the components perpendicular to the shock normal scale with the compression ratio, $r$, as $B_{post}/B_{pre} \propto r$ (with equality in perpendicular shocks). For these weak shocks, the compression ratio is close to unity, resulting in minimal amplification of the magnetic field in the perpendicular direction.
 In the upper sub-plots of each panel we show the Mach number distribution of those shocks that are in full or partial contact with the CR jet material (shown in blue). We additionally identify which shocked cells are in contact with the CR jet material and show the corresponding Mach number distribution (shown purple). Shocks generated early on in the merger evolution (see upper left panel) are stronger with $\mathcal{M}\sim2$--3 than those interacting with the jet material at later times, $\tau\gtrsim 5$ Gyr, with $\mathcal{M}\sim1$--2.

 We will now discuss the importance of distinguishing between the early and late evolutionary phases in understanding these contributions:
 \begin{itemize}
     \item[i)] \textit{Early merger phase and early burst}: During this phase where the shocks are stronger, the jet material is not volume filling because the lobes are in the process of being completely disrupted. Therefore, merger shocks during this epoch would only partially interact with the jet material. For example, in the upper panels of Fig.~\ref{fig:relics} we can see that $\lesssim 1/2$ of the shock interacts with the jet material. Furthermore, in this example, the Mach number distribution corresponding to the part of the shock that interacts with the jet material has a mean $\mathcal{M}\sim1.4$. It is still not understood whether DSA would work at such weak shocks (see point below).
     \item[ii)]\textit{Late merger phase and old lobe material}: During this phase, the jet material has already permeated the central region of the cluster (see also Figs.~\ref{fig:multipanel_evolution}--~\ref{fig:multipanel_evolution_theta10}). Hence, it is more likely for merger shocks to fully interact with the CR jet material. This in principle would mean that the fossil population of electrons could participate in the DSA and account for the whole area of the radio shock. But here the generated shocks are already weak. Results from particle-in-cell and hybrid-kinetic simulations suggest that there is a critical Mach number, $\mathcal{M}_{cr}\sim 2.3$, below which thermal electrons cannot participate in DSA \citep[][]{2021ApJ...915...18H} because such shocks do not develop shock surface ripples \citep[][]{2024A&A...684A.129B}. For the case of fossil electrons, this is yet to be understood. Pre-energized electrons and/or turbulence could influence electron acceleration in subcritical shocks, such as the ones shown in Fig~\ref{fig:relics}. Nevertheless, \citealt{2022ApJ...925...88H} argues that pre-energized electrons in the pre-shock region alone would not resolve the issue of electron pre-acceleration at sub-critical ICM shocks. 
 \end{itemize}
 Therefore we conclude that merger shocks that could be observed in radio and associated with a single outburst from a central AGN should be scarce based on the low volume fraction of the shocked CR jet material and the associated low mean Mach number in the majority of our runs (see Section~\ref{sec:evolution_jet_material}). A combination of multiple jet outbursts and/or off-center radio galaxies and/or multiple mergers would enhance the probability of finding these merger shocks lightening up in the radio band through shock re-acceleration. Another possibility would be to have an earlier outburst that has already spread the CR jet material around the center of the cluster $\sim 1$ Gyr before the merger.

%

\subsection{Contribution to radio halo emission?}
\label{sec:halos}

In this section, we investigate whether the CR jet material could be efficiently accelerated by turbulence and become luminous in the radio band. We analyze average quantities related to turbulence and energy losses throughout the temporal evolution of the CR jet material, as indicated by the jet tracers. This is of particular interest for the formation of radio halos or mega halos whose observed statistical properties seem to indicate that turbulent re-acceleration is responsible for the observed radio emission \citep[e.g.,][]{2001MNRAS.320..365B, 2001ApJ...557..560P,2003ApJ...584..190F,2024ApJ...961...15N}.

The presence of turbulence and magnetic field amplification via small-scale dynamos is suggested to be ubiquitous within galaxy clusters, as indicated by cosmological MHD simulations \citep[e.g.,][]{2018MNRAS.474.1672V,2019MNRAS.486..623D,2022ApJ...933..131S}.  
Turbulent energy can be transferred into CRs through
stochastic wave-particle interactions.
The two turbulent re-acceleration mechanisms that are broadly discussed in the literature are: i) Transit Time Damping (TTD) model which considers the resonant
interaction with the fast magneto-sonic compressive modes of the turbulence \citep[e.g.,][]{brunetti2007,2015ApJ...800...60M,2017MNRAS.465.4800P}, and ii) the super-Alfv\'enic non-resonant acceleration model which considers stochastic diffusion across magnetic reconnection and dynamo regions and interaction with the super-Alfv\'enic solenoidal turbulence \citep[][]{2016MNRAS.458.2584B}.
In these cases, the momentum diffusion coefficient becomes hard-sphere
type, that is $D_{pp} \propto p^2$.
Previous MHD simulations studies show that the turbulent velocity solenoidal modes
typically dominate over compressive modes
in the ICM \citep[e.g.,][]{2015ApJ...800...60M,2015ApJ...810...93P,2017MNRAS.464..210V}. 
We compute the turbulent kinetic energy flux as

\begin{figure}
    \centering

    \includegraphics[width=\columnwidth]{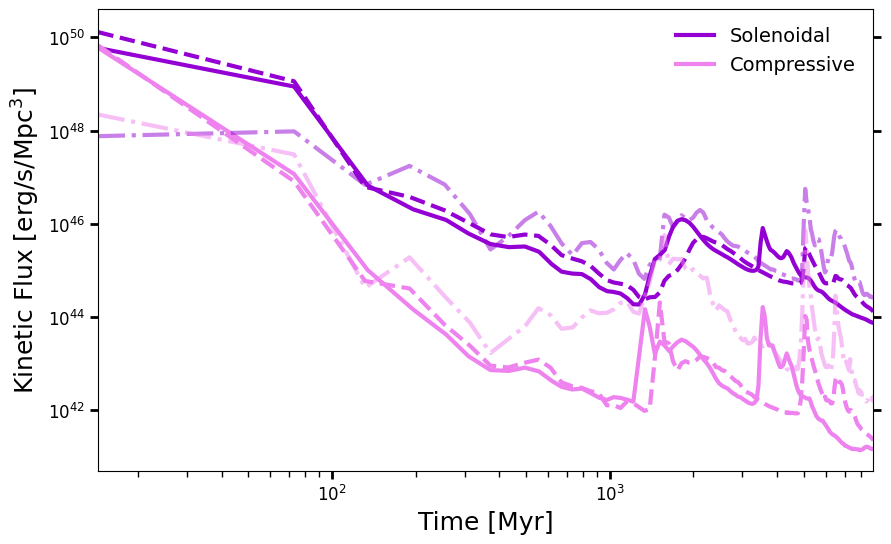}
    \caption{Volume average of the turbulent energy flux per unit volume for solenoidal (purple) and compressive (pink) modes as a function of time. We show the evolution for the R=1:2 (solid line) and R=1:5 (dashed line) runs with $\theta=20^{\circ}$, $t_{\mathrm{jet,0}}=50$ Myr and the jet orientation initialized in the $x$-direction. The dash-dotted lines show the flux computed with the solenoidal and compressive turbulent velocities using the Helmholtz-Hodge decomposition method for the R=1:2 run.}
    \label{fig:kinetic_flux}
\end{figure}

\begin{equation}\label{eq:flux}
    F_{\mathrm{turb}} = \frac{1}{2} \frac{\rho \, \delta V^{3}(L_0)}{L_0} (\Delta x)^3,
\end{equation}
where $\rho$ is the density, $\delta V$ is the velocity fluctuation measured within the scale $L_0$ and $\Delta x$ is the typical cell size. Note that $\delta V^{3}/L_0$ is the specific energy rate. 
We use two methods to keep track of these the solenoidal and compressive turbulent velocity modes in the cells where the jet material is located in our simulations. In the first method,
we used the gas vorticity measured by the jet tracers to estimate the local level of solenoidal turbulence, $\delta V_{sol}=\mid \nabla \times \mathbf{v} \mid L_0$, and compressive turbulence, $\delta V_{com}=\mid \nabla \cdot \mathbf{v} \mid L_0$. We used as reference scale the fixed cell scale used to compute either the divergence or vorticity\footnote{We output the volume of the Voronoi cells and for the sake of simplicity we assumed that these are spherical to extract a typical value of $\Delta x$. Note that, in this case, $L_0$ is not rigurously within the inertial range.}. The kinetic turbulent velocity spectrum could be sensitive to the dynamical state of the cluster and to the numerical scheme. There have been previous numerical studies reporting either different slopes for the solenoidal and compressive components \citep[e.g.,][]{2014ApJ...782...21M,2015ApJ...800...60M,2017MNRAS.464..210V} or roughly same slopes agreeing with the Kolmogorov prediction \citep[][]{Valles_Perez_2021b}. We assume the latter is true for simplicity. Under this assumption, $F_{\mathrm{turb}}$ is insensitive to the specific choice of $L_0$ as long as it is in the inertial range. 
In the second method, we use the vortex-p code developed by \citealt{Valles_Perez_2024} to find the solenoidal and compressive components performing a Helmholtz-Hodge decomposition in Fourier space and the multi-filtering technique described in \citealt{Valles_Perez_2021a, Valles_Perez_2021b}.\footnote{We use the public GitHub repository \url{https://github.com/dvallesp/vortex-p}}
This algorithm iteratively constrains the turbulence outer scale, $L_0$, at each point in the mesh. In this way, the turbulent velocity field, $\delta V (L_0)$, is computed without having to fix the filtering length. We refer the reader to Appendix \ref{app:vortex} for further details.

In Fig.~\ref{fig:kinetic_flux}, we present the time evolution of the volume-averaged turbulent energy flux per unit volume, for both solenoidal and compressive modes, at the location of the CR material. We show the evolution for runs with $\theta=20^{\circ}$, $t_{\mathrm{jet,0}}=50$ Myr, and jets oriented in the $x$-direction for both mass ratios. We show the results from the two methods used to separate the solenoidal and compressive turbulent components.
Similar time evolution patterns occur in other runs, so they are not shown here. Notably, apart from the initial burst and the first core passage (indicated by the first spike around $\sim 1$ Gyr in Fig.~\ref{fig:kinetic_flux}), the solenoidal component dominates throughout the entire period where jet material mixes with the ICM. This is true independently of the method used to separate the solenoidal and compressive turbulent components.
In the remainder of this paper, we use the Helmholtz-Hodge decomposition method to compute the solenoidal turbulent velocity.
In the following we will explore the efficiency of the turbulent acceleration particle mechanism by these solenoidal modes.

CRe typically undergo energy losses in the cluster environment due to inverse Compton (IC), synchrotron, Bremsstrahlung, and Coulomb interactions. For re-acceleration processes to be effective, they must occur more rapidly than these energy losses. We first calculate the cooling timescales using volume-averaged fluid quantities traced by the jet material. The synchrotron and IC losses can be modelled as \citep[][]{1979rpa..book.....R}
\begin{equation}
    \left| \frac{dp}{dt} \right|_{\mathrm{Sync+IC}} = \frac{4}{9} r_0^2 \beta^2 \gamma^2 
    \left[ B^2 + B_{\mathrm{CMB}}^{2} (1+z)^4\right],
\end{equation}
where $p$ is the momentum, $r_0=q_e^2/(m_e c^2)$ is the electron radius, $\beta^2=1-\gamma^2$, $\gamma=p/(m_e c)-1$ and $B_{\mathrm{CMB}}$ is the IC equivalent magnetic field due to the cosmic background radiation (CMB). The Coulomb losses can be estimated \citep[][]{2002cra..book.....S} as
\begin{equation}\label{eq:Coul}
     \left| \frac{dp}{dt} \right|_{\mathrm{Coul}} =
     \frac{4\pi r_0^2 n_{th} m_e c^2}{\beta_e} \ln \Lambda,
\end{equation}
where $n_{th}$ is the thermal number density, 
\begin{equation}\label{eq:sync}
    \ln \Lambda = 37.8 + \log \left( \frac{T_{[K]}}{10^8 \, \mathrm{K}} \left(\frac{n_{th}}{10^3 \, \mathrm{cm}^{-3}}\right)^{-1/2} \right),
\end{equation}
\begin{figure}
    \centering

    \includegraphics[width=\columnwidth]{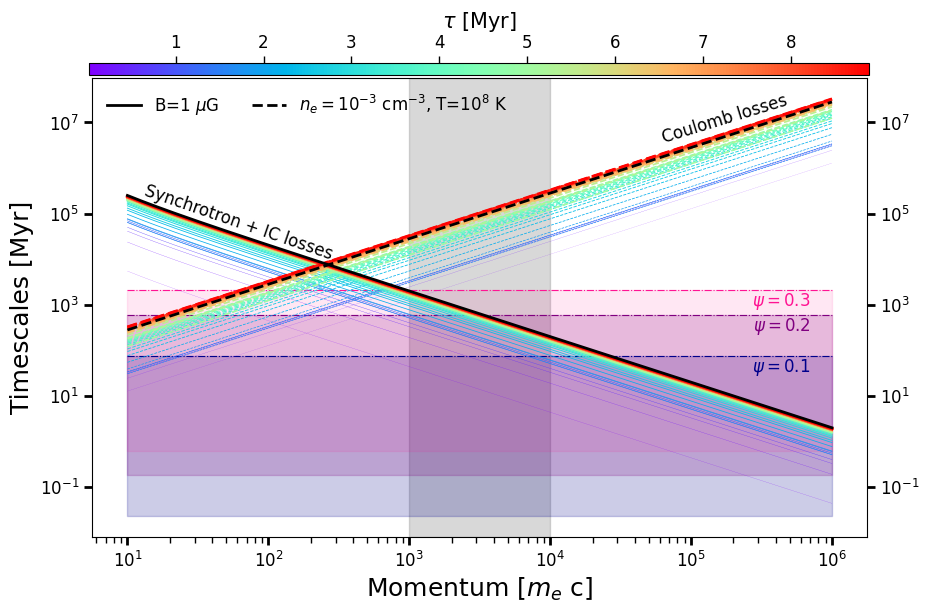}
    \includegraphics[width=\columnwidth]{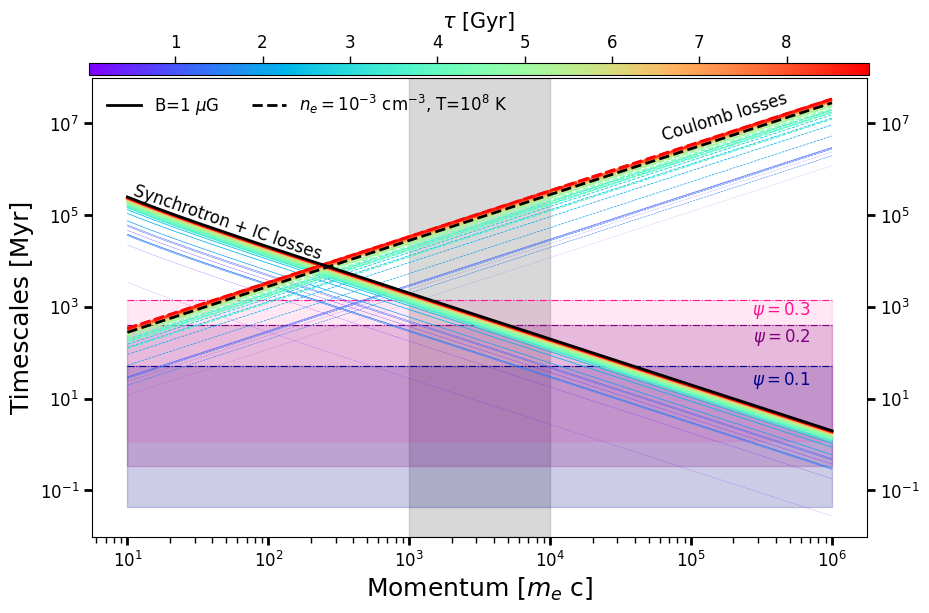}

    \caption{Timescales of synchrotron + IC losses (solid lines), Coulomb losses (dashed lines) and super-Alfv\`enic turbulent (re-)acceleration (blue dot-dashed lines) as a function of momentum. We show the R=1:5 (upper panel) and R=1:2 (lower panel) cases with $\theta=20^{\circ}$, the jet initialized at $t_{\mathrm{jet,0}}=50$ Myr and in the $x$-direction. The losses lines are colored according to the evolution after the launching time. The acceleration times computed with Eq.~\ref{eq:Dpp} for different values of $\psi$ are shown in the horizontal shaded areas. The extent of each shaded area corresponds to the minimum and maximum estimated $t_{\mathrm{acc}}$ throughout the simulation.}
    \label{fig:timescales} 
\end{figure}
is the Coulomb logarithm at $T>4\times 10^5$ K \citep[][]{1999ApJ...520..529S} and $T_{[K]}$ is the temperature in units of K. In Fig.~\ref{fig:timescales} we show the typical timescales as a function of momentum. 
We illustrate the time evolution of energy losses using differently colored lines, as calculated from Eqs.~\ref{eq:Coul}--\ref{eq:sync}. These calculations are based on the fluid average quantities from the jet material at each timestep. We omit the Bremsstrahlung cooling timescales as they are generally larger than those of interest in our study. 
The cooling timescale for the radio-emitting electrons with $p= 10^3$–$10^4 m_e c$\footnote{The synchrotron emission peaks at $\nu_{\mathrm{peak}}=0.29 \nu_c = (3eB/4\pi m_ec) \gamma^2=1.2597 \gamma^2 B_{[\mu G]}$ Hz, where $\nu_c$ is the critical frequency, and $B_{[\mu G]}$ is the magnetic field strength in units of $\mu$G. For example, for a typical value of $\sim3.5 \mu$G, the emission at $\sim 100$ MHz is due to electrons with $\gamma\sim5\times10^3$.}
(see gray area in Fig.~\ref{fig:timescales})  is about 0.1–1 Gyr for a typical magnetic field strength of a few $\mu$G. To aid in visualization, we have included the expected synchrotron + IC (assuming $z=0.018$) and Coulomb cooling times for typical ICM values, represented by black lines. We show two representative cases with $\theta=20^{\circ}$ and the jet initiated at $t_{\mathrm{jet,0}}=50$ Myr in the $x$-direction. The upper panel depicts the mass ratio R=1:5, while the lower panel shows R=1:2. We also overplot the re-acceleration timescale
\begin{equation}\label{eq:tacc}
    t_{\mathrm{acc}} =  \frac{p^2}{4 D_{pp}},
\end{equation}
where $D_{pp}$ is the re-acceleration diffusion coefficient in the particle momentum space in the isotropic Fokker-Planck equation \citep[][]{2002cra..book.....S}. The coefficient $D_{pp}$ depends on the particles mean free path. 
In our simulations, the turbulent motions are mainly super-Alfv\`enic \footnote{The mean Alfv\`en Mach number in regions with CR jet material is in the range of $\sim 4$-9 throughout the whole evolution of the merger.}. In this case,
the interaction is dominated by the largest reconnection regions \citep[][]{2016MNRAS.458.2584B}, meaning that the maximum effective mean free path is limited to the Alfv\`en scale, $l_A=L_{0} \mathcal{M}_A^{-3}$ (Kolmogorov scaling), where $\mathcal{M}_A=\delta V/v_A$ is the turbulent Alfv\`en Mach number and $v_A$ is the Alfv\`en speed. Following \cite{2016MNRAS.458.2584B}, in this mechanism one can write the diffusion coefficient as
\begin{equation}\label{eq:Dpp}
    D_{pp} \simeq 3 \sqrt{\frac{5}{6}} \frac{c_s^2}{c}
    \frac{\sqrt{\beta_{p}}}{L_{0}} \mathcal{M}_{t}^{3} \psi^{-3} p^2
\end{equation}
where $\psi$ defines an average effective mean free path (mfp) as a fraction of $l_{A}$, that is $\lambda_{\mathrm{mfp}}= \psi l_A$, where $\psi<1$ \citep[see][for studies where Eq.~\ref{eq:Dpp} has also been computed from MHD simulations]{2020PhRvL.124e1101B,2024ApJ...961...15N} and $\mathcal{M}_t=\delta V /c_s$ is the turbulent sonic Mach number. 
\begin{figure*}
    \centering
    \includegraphics[width=\textwidth]{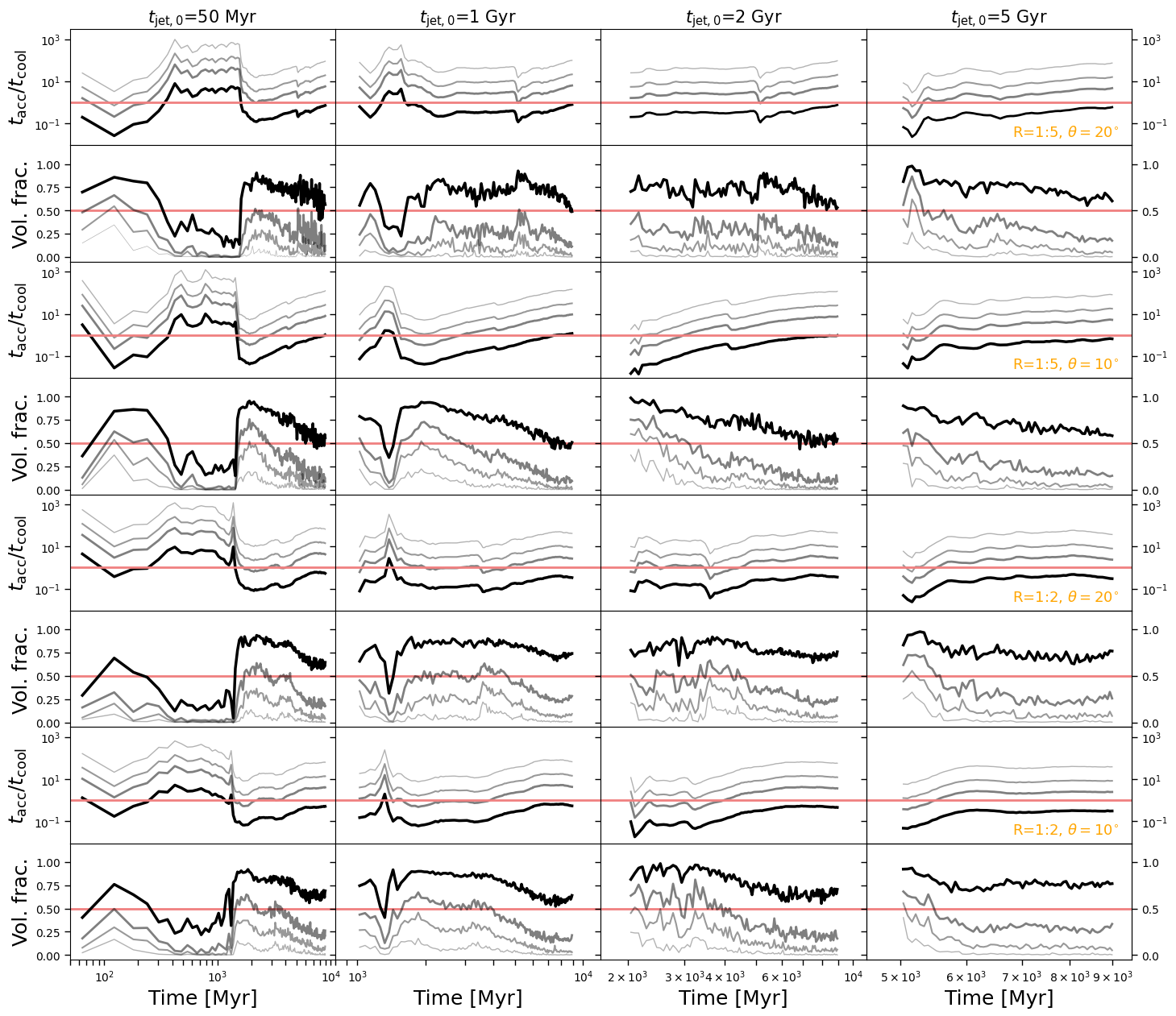}
    \caption{Acceleration and cooling time ratio, $t_{\textrm{acc}}/t_{\mathrm{cool}}$ (1st, 3rd, 5th and 7th rows), and volume fraction where $t_{\textrm{acc}}/t_{\mathrm{cool}}<1$ (2nd, 4th, 6th and 8th rows) as a function of time for all our simulations at a fixed momentum value of $5\times 10^3m_e c$ (where electrons in a $\mu$G magnetic field will contribute to the $\sim100$ MHz emission). Similar to Fig~\ref{fig:multipanel_evolution}, each column shows the evolution for runs with different initial jet ignition times, $t_{\mathrm{jet,0}}$. We only show runs with an initial jet direction in the $x$-direction. 
    Larger line widths and smaller opacity lines refer to a smaller parameter value $\psi$. The horizontal gray (coral) solid lines are a reference for $t_{\textrm{acc}}/t_{\mathrm{cool}}=1$ (50\% of the volume fraction).}
    \label{fig:tfrac} 
\end{figure*}

In Fig.~\ref{fig:timescales}, we show the acceleration timescale for different values of $\psi$ using Eqs.~\ref{eq:tacc}-\ref{eq:Dpp} (see horizontal shaded areas). Similar to the energy losses, each line in this plot would correspond to a different timestep. For purposes of better visualization, we show only shaded areas encompassing the minimum and maximum estimated $t_{\mathrm{acc}}$ within the jet material at a fixed $\psi$ value. 
It has been previously argued that $\psi$ should satisfy $0.1 \lesssim \psi \lesssim 0.5$ because of the effect of particle scattering due to mirroring in a super-Alfv\`enic flow \citep[][]{2016MNRAS.458.2584B}. In Fig.~\ref{fig:timescales} we show the parameter values $\psi=0.1$, $\psi=0.2$, and $\psi=0.3$. Increasing values of $\psi$ lead to longer acceleration timescales. Nevertheless, it is only when $t_{\mathrm{acc}}<t_{\mathrm{cool}}$ that this turbulent re-acceleration mechanism can efficiently re-accelerate CRe that radiate synchrotron emission. Our analysis using jet tracers shows that during the evolution and mixing of the jet material within the ICM, the diffusion coefficient $D_{pp}$ varies over time. Specifically, the acceleration timescale $t_{\mathrm{acc}}$ spans from approximately 0.02 Myr to tens of Myr for $\psi=0.1$, 0.2 Myr to hundreds of Myr for $\psi=0.2$, and 0.6 Myr to a few Gyr for $\psi=0.3$. 
We see that these results are almost independent of the mass ratio of the merger (compare upper and lower panels in Fig.~\ref{fig:timescales}). 

A better way to visualize the periods when the turbulent acceleration energy gain could be more efficient than the cooling energy losses is shown in Fig.~\ref{fig:tfrac}. In this figure, we show the median of the $t_{\mathrm{acc}}/t_{cool}$ distribution\footnote{We note that the $t_{\mathrm{acc}}/t_{cool}$ distribution is a highly positive skewed function and therefore, we show the median evolution.} and the volume fraction for cells where $t_{\textrm{acc}}/t_{cool}<1$ as a function of time for all our simulations. The initial direction of the jet does not significantly alter the evolutionary trends shown in Fig.~\ref{fig:tfrac} so we only show the results from those simulations where the jet was initialized in the $x$-direction. 
We use Eq.~\ref{eq:Dpp} to calculate $t_{\mathrm{acc}}$ and in this case we show the results for the parameter values $\psi=0.1, 0.2, 0.3$ and 0.5, with a fixed momentum value of $5\times 10^3m_e c$ (see footnote 10). The typical pattern for the fraction of these timescales is characterized by an increment at the time of the first core passage, followed by a gradual decline. In the $\theta=10^{\circ}$, $t_{\mathrm{jet,0}}=1$ Gyr and $t_{\mathrm{jet,0}}=2$ Gyr cases (see second and third columns), $t_{\textrm{acc}}/t_{cool}$ slowly rises in the last $\sim5$ Gyr of the simulations. 
Notably, one can see that an increase in $t_{\mathrm{acc}}/t_{cool}$ during the first core passage comes along with a decrease in the volume fraction, which can drop to as low as $\sim$0.1\% (see first column of Fig~\ref{fig:tfrac}).
This is particularly expected in the $t_{\mathrm{jet,0}}=50$ Myr cases, where the first core passage can quickly displace the CR jet material to the cluster's outskirts. This would make the radio emission less volume filling. 
However, we see that, shortly after $\sim0.5$ Gyr, the volume fraction could easily increase to $\gtrsim 50$\%.
There are several things to note: 1) This re-acceleration mechanism appears to be effective only during specific intervals; 
2) Parameters $\psi\sim0.2$--0.3 result in periods typically lasting $1$--2 Gyr where $t_{\textrm{acc}}/t_{cool}<1$. These conditions correspond to volume fractions ranging from $\sim 20$\% to $\sim 70$\%, indicating the possibility of volume filling emission. The typical mean values of the magnetic field strength in the regions containing CR jet material, $\sim1$--10 $\mu$G during the evolution of the merger, and the selected momentum imply radio emission at 100s of MHz frequencies.
Specifically, $\psi\sim0.3$ provides a more conservative estimate, ensuring a significant volume filling emission that may generate diffuse radio emission with a steep spectrum. Conversely, $\psi=0.1$ is too efficient, maintaining $t_{\textrm{acc}}/t_{cool}<1$ throughout most of the simulation and achieving volume fractions exceeding 70\%. 
More importantly, a non-significant volume fraction with $t_{\textrm{acc}}<10$ Myr would generate too much power;
3) The initial jet ignition time, $t_{\mathrm{jet,0}}$, in relation to the evolution of the merger is critical. Particularly, if the jet burst occurs briefly before the first core passage, there will be a larger displacement of the CR jet material and the compressive turbulent energy flux becomes relevant (see also Fig.~\ref{fig:kinetic_flux}). This results in longer re-acceleration timescales than the cooling timescales, thus reducing the efficiency of the turbulent re-acceleration mechanism.

It is not entirely understood whether there is a specific onset of turbulent re-acceleration related to minor and major merger events and accretion flows \citep[e.g.,][]{2016A&A...593A..81C} or multiple onsets. Our findings suggest that the super-Alfv\`enic re-acceleration mechanism exhibits periodic efficiency, particularly noticeable after $\sim$0.5 Gyr following the second core passage. This is evidenced by an increase in the volume fraction to over 50\% in most of our simulations.

While studying the evolution of CRe with a Fokker-Planck solver would complement our current conclusions, our findings suggest that a single merger event producing sloshing motions coupled with AGN activity effectively facilitates super-Alfv\`enic re-acceleration mechanism. In reality, galaxy clusters experience multiple minor and major mergers over their lifetimes, which would only generate more turbulence and/or sustain it for longer. 
In this case, we can conclude that turbulent re-acceleration would be an even more plausible mechanism to explain the extended diffuse radio emission in galaxy clusters. We leave the comparison of these results with other turbulent re-acceleration and/or secondaries models for future work.

Our study further supports the important role of jets and AGN in providing a pool of seed electrons that can be re-accelerated. 
%

\section{Summary and conclusions}
\label{sec:conclusions}

We conducted a suite of binary galaxy cluster merger MHD simulations using the moving-mesh code AREPO, including a bi-directional jet at the core of the primary galaxy cluster. We varied the initial mass ratio (R=1:2 and R=1:5), impact parameter ($\theta=10^{\circ}$ and $\theta=20^{\circ}$), jet orientation ($x$-, $y$-, $z$-direction), and jet ignition time ($t_{\mathrm{jet},0}=0.05$, 1, 2, and 5 Gyr). Additionally, our simulations incorporated a CR component in the jet model, treated using the two-fluid approximation. The CR energy density and CR pressure terms are included in the MHD treatment of the AREPO code. In this first parameter space study, the CR fluid is mainly advected and we did not include Alfv\`en cooling, spatial diffusion and streaming terms. This study focuses on the interaction between the jet and the ICM, particularly examining the lifetime and spatial distribution of CR material during merger evolution. We also analyzed the effects of merger-induced shocks on the CR material. Our main findings can be summarized as follows:

\begin{itemize}
    \item[i)] \textit{Spatial distribution}: Any initial merger and/or jet configuration leads to a final CR spatial distribution concentrated towards the center of the main cluster with a roundish morphology in $\sim5$--6 Gyr. The maximum spatial extension of the CR material is $\sim1$--2 Mpc.
    
     \item[ii)]\textit{The role of the mass ratio and impact parameter}: The mass ratio parameter plays a more important role than the impact parameter for the final maximum spread of the CR injected by the jet. A larger mass ratio results in stronger shocks (with maximum mean Mach numbers of $\sim$2.5 and $\sim$3.5 for $R=1:5$ and $R=1:2$, respectively) and a higher central entropy of the cluster.
     
    \item[iii)] \textit{Timing and initial direction of the outburst}:  The later the AGN burst occurs during the merger, the less interaction there is between the CR jet material and merger shocks. While the initial direction of the jet and the initial time of the outburst do not significantly impact the late evolution of the CR jet material, they are relevant for replicating specific radio source morphologies in the early phase of the jet.

    \item[iv)] \textit{Mixing}: An ongoing binary merger leads to rather quick mixing of the CR jet material into the ICM, lasting around 1.5 Gyr.

    \item[v)] \textit{Contribution to the radio relic emission}: Merger-induced shocks interact with the CR jet material; however, it appears unlikely that this material significantly contributes to the radio relic emission associated with these shocks. In the early phases of the merger, only a small volume fraction of the CR jet material is shocked. Later, as the merger progresses, the shocks are weaker ($\mathcal{M}\lesssim 2$), making it challenging for fossil electrons to participate in DSA. 

    \item[vi)] \textit{Contribution to the radio halo emission}: Due to the large radii that the CR jet material can reach and the amount of (solenoidal) cluster turbulence, it is plausible for the CR jet material to be re-accelerated.
    We examined the cooling and re-acceleration timescales in the case of super-Alfv\`enic turbulent re-acceleration and found that this mechanism could be periodically efficient, leading to significant volume fractions where $t_{\textrm{acc}}/t_{cool}<1$.
\end{itemize}

While the primary focus of this work is on the late-phase propagation of jet CRs throughout the main galaxy cluster, we also discussed the initial phase, where we found a diversity of morphologies including WAT-like and X-shaped-like sources. In our simulations, the emergence of WAT-like sources is influenced more by the timing of the initial radio outburst and the subsequent merger, rather than the merger mass ratio. Specifically, these sources typically develop within 250 Myr after the first core passage and $\sim$1.75 Gyr after the initial outburst.
On the other hand, we spot a few X-shaped-like sources in our simulations. These emerge only with the R=1:2 mass ratio and a rather later initial burst happening after the first core passage. Overall, our simulations highlight the relevance of the cluster environment in shaping various radio sources.

We showed that the late propagation of the CR jet material fills up the core region of the cluster which could provide seed electrons for radio mini-halos as originally proposed by \cite{2013ApJ...762...78Z} and possibly even radio halo-like emission. %
We explored the super-Alfv\`enic turbulent re-acceleration model proposed by \cite{2016MNRAS.458.2584B} and showed that the associated re-acceleration time can be shorter than the cooling time over a significant volume at the center of the cluster and for long periods of time. While this holds true for all our simulations, the periods during which this mechanism is efficient vary from case to case. This highlights the importance of the interaction between the cluster dynamics and the jet.
The acceleration efficiency depends on the mean free path of a CR particle, which is expressed in terms of the parameter $\psi$. 
We suggest that a parameter value of $\psi\sim0.3$ is enough to re-accelerate CRe effectively during periods of $\sim1$--2 Gyr and produce a volume filling emission, $\gtrsim 50$\%, which would potentially be observable at MHz frequencies given the typical $\sim\mu$G values found within the central region of the cluster.
 To more accurately model the energetic properties of the CR that can produce radio emission, we should make use of a Fokker-Planck solver and tracer particles \citep[see e.g.,][for examples using cosmological simulations]{2023A&A...669A..50V,2023MNRAS.519..548B}. Therefore, future work is required to  confirm whether this mechanism could fully explain extended radio emission. 
Furthermore, it is clear that other factors could come into play here, such as multiple mergers that generate more turbulence and/or multiple AGN bursts that provide more seed electrons and/or multiple onsets of turbulent re-acceleration and/or the action of other turbulence re-acceleration processes and/or the generation of secondary electrons.

The same limitations apply for the case of radio relics. We do not model the DSA acceleration at shocks and follow the energy losses \citep[e.g.,][]{dominguezfernandez2020morphology,dominguez2024patchiness}. Nevertheless, we do not expect that a more complicated CR modelling will change our main conclusions regarding how challenging would it be to get a substantial contribution from the interaction of a central AGN with merger shocks to the radio relic emission. In this case, our conclusions rely mostly on the properties of the merger, strength of the shock, and timing of the jet ignition. We believe that a combination of multiple jet outbursts
and/or off-center radio galaxies and/or multiple mergers
would enhance the probability of finding these merger
shocks lightening up in the radio band through shock
re-acceleration.

In future work we will include the modelling of synchrotron emission following the ageing of CRe, study the role of different AGN cycles, different jet radio powers, and off-center and multiple AGNs. Finally, we also recognize that in reality, Alfv\`en cooling, diffusion and/or streaming would smooth the CR distribution, potentially affecting the observed radio emission \citep[][]{2021Galax...9...91Z}. This aspect merits further investigation.

%



\section*{Acknowledgments}
%
%
We acknowledge useful discussions with Elizabeth Blanton and Ewan O'Sullivan. We would also like to acknowledge the main developer of the CR module in AREPO, Christoph Pfrommer. We would like to thank David Vall\'es-P\'erez for his support on the use of the vortex-p code. We would also like to thank the anonymous reviewer for useful comments that increased the quality of this manuscript.
The simulations
presented in this work made use of computational resources on the
Canon cluster at Harvard University. P. Dom\'inguez-Fern\'andez acknowledges the Future Faculty
Leaders Fellowship at the Center for Astrophysics, Harvard-Smithsonian. R. Weinberger acknowledges funding of a Leibniz Junior Research Group (project number J131/2022).
%





\software{ The source codes used for
the simulations of this study, AREPO \citep{2010MNRAS.401..791S,2011MNRAS.418.1392P} is freely available
on \url{https://arepo-code.org/}. The turbulent velocity decomposition was done using vortex-p \citep{Valles_Perez_2021a, Valles_Perez_2021b}, which is freely available on \url{https://github.com/dvallesp/vortex-p}. The main tools for our analysis are:
        python \citep{van1995python}, matplotlib \citep{Hunter:2007}, numpy \citep{harris2020array} and  
          the yt analysis toolkit \citep{2011ApJS..192....9T} freely available at \url{https://matplotlib.org/}, \url{https://www.numpy.org} and \url{https://yt-project.org/}. 
          }

\section*{Data Availability Statement}

%
The data underlying this article includes 48 low-resolution simulations and 5 high-resolution simulations, amounting to several TB in total. Due to its size, it cannot be deposited in public archives. However, the authors are willing to share portions of the simulations upon reasonable request to the corresponding author.


\appendix

\section{Projection maps of the all the runs}
\label{app:theta10}

\begin{figure*}
    \centering
    \includegraphics[width=0.85\textwidth]{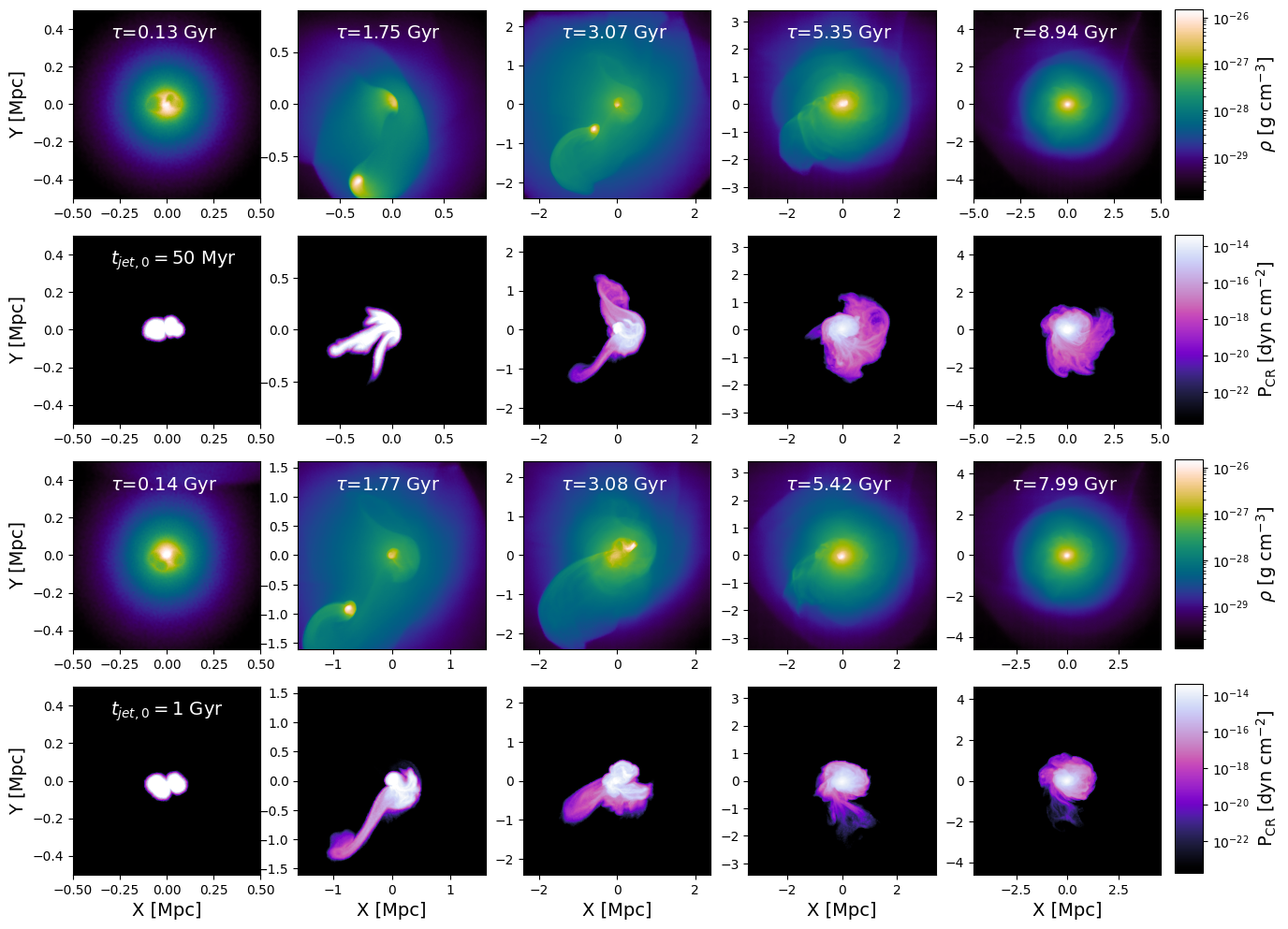}
    \includegraphics[width=0.85\textwidth]{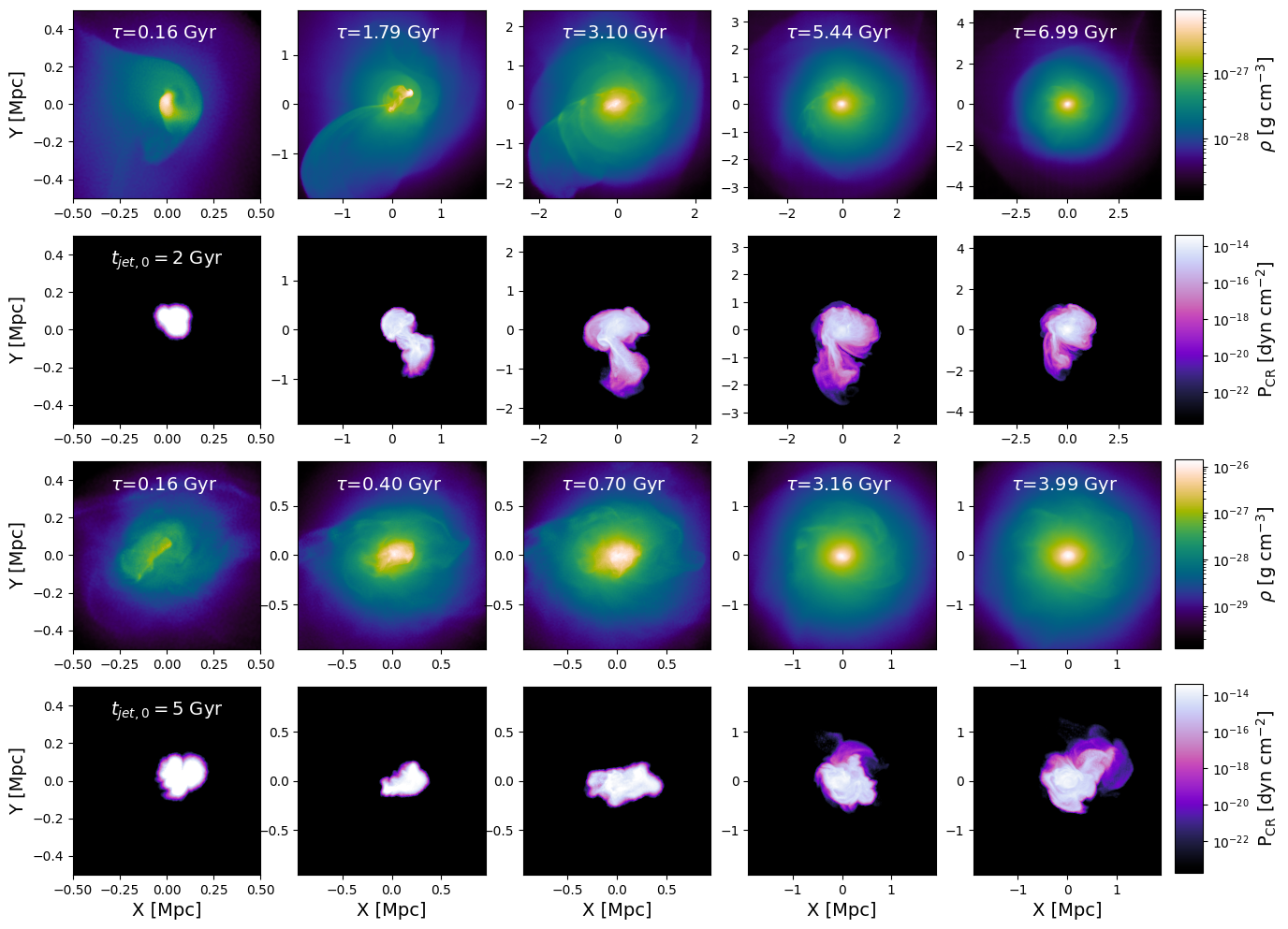}
    \caption{Density and CR pressure projection maps in the $z$-direction. Case $R=1:2$, $\theta=20^{\circ}$ and starting jet activity after 50 Myr and 1 Gyr (upper panels), and 2 Gyr and 5 Gyr (lower panels). Note that Note that $\tau=t - t_{\mathrm{jet},0}$ and that the times shown are different for the 5 Gyr case.}
    \label{fig:maps_r0p5}
\end{figure*}

\begin{figure*}
    \centering
    \includegraphics[width=0.84\textwidth]{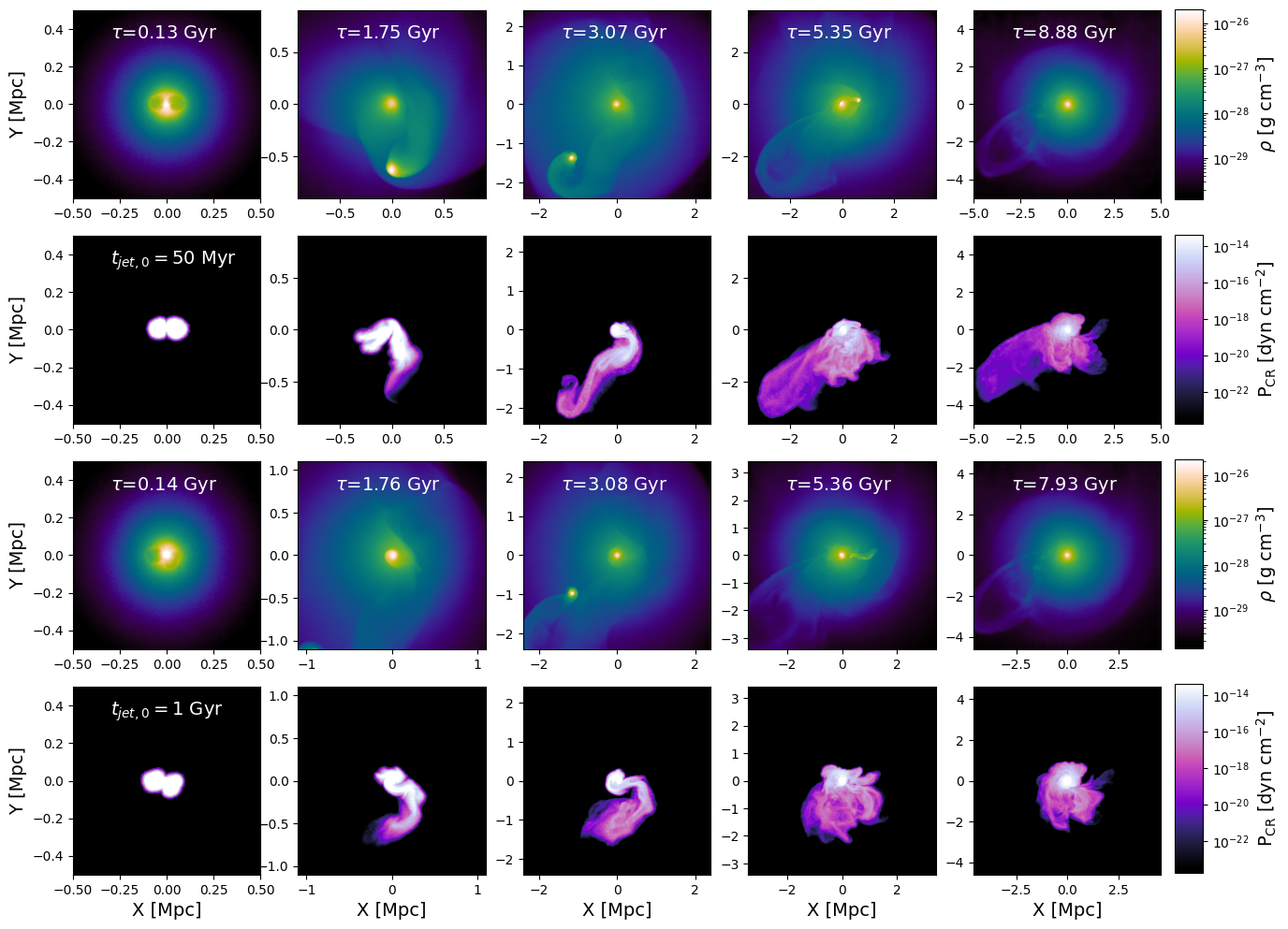}
    \includegraphics[width=0.84\textwidth]{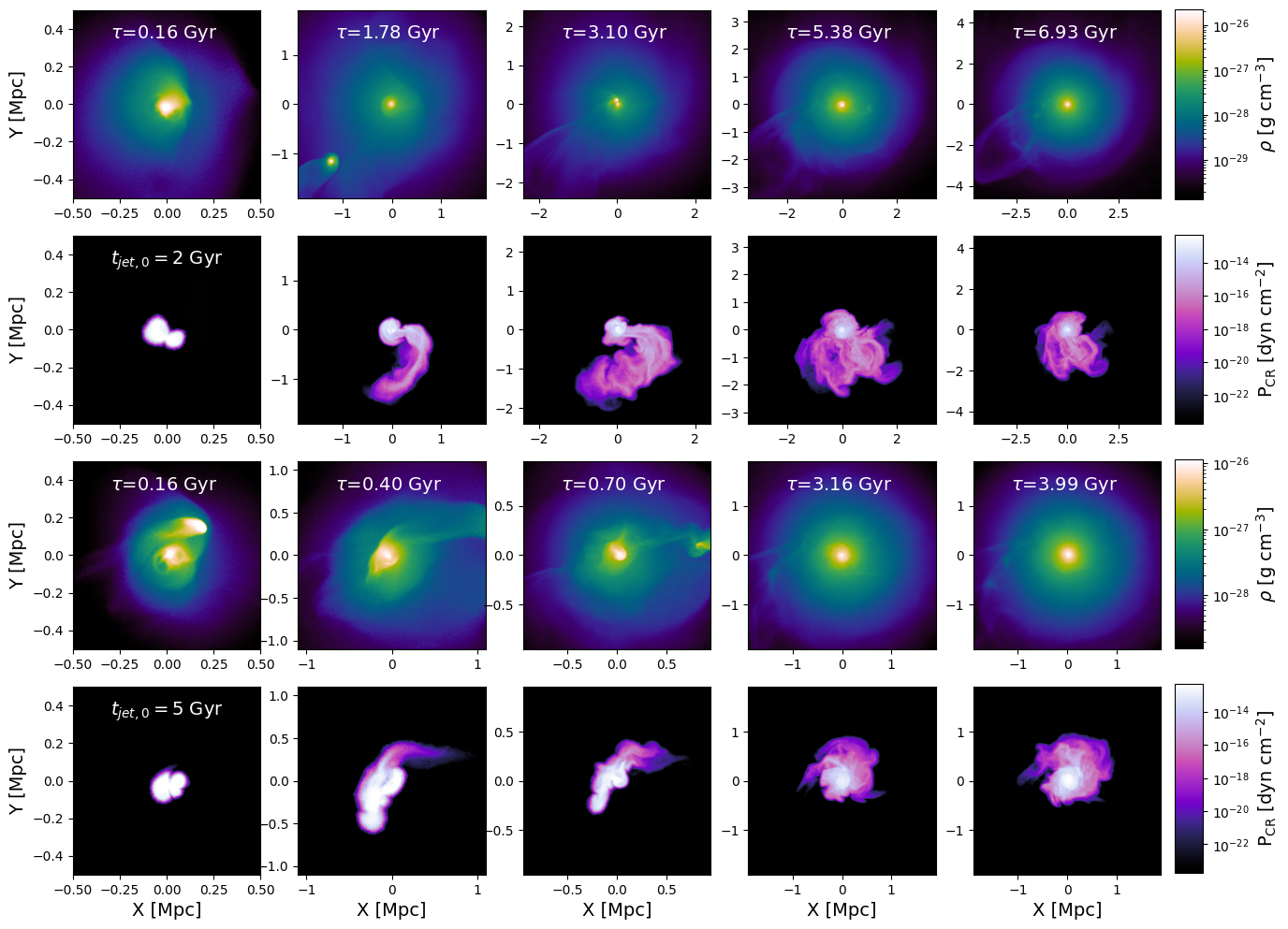}
    \caption{Case $R=1:5$, $\theta=20^{\circ}$ and starting jet activity after 50 Myr and 1 Gyr (upper panels), and 2 Gyr and 5 Gyr (lower panels). Note that Note that $\tau=t - t_{\mathrm{jet},0}$ and that the times shown are different for the 5 Gyr case.}
    \label{fig:maps_r0p2}
\end{figure*}

\begin{figure*}
    \centering

    \includegraphics[width=0.85\textwidth]{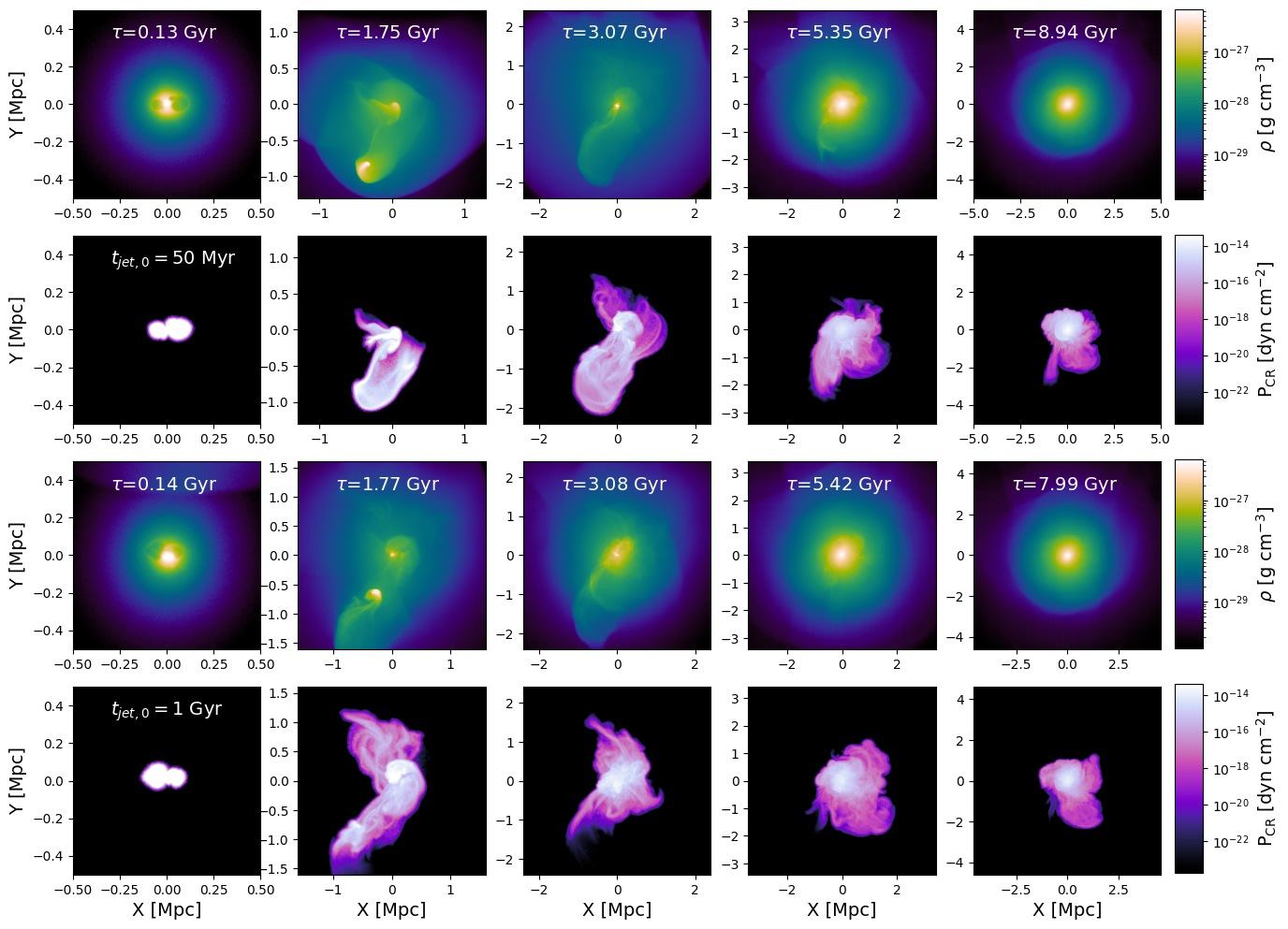}
    \includegraphics[width=0.85\textwidth]{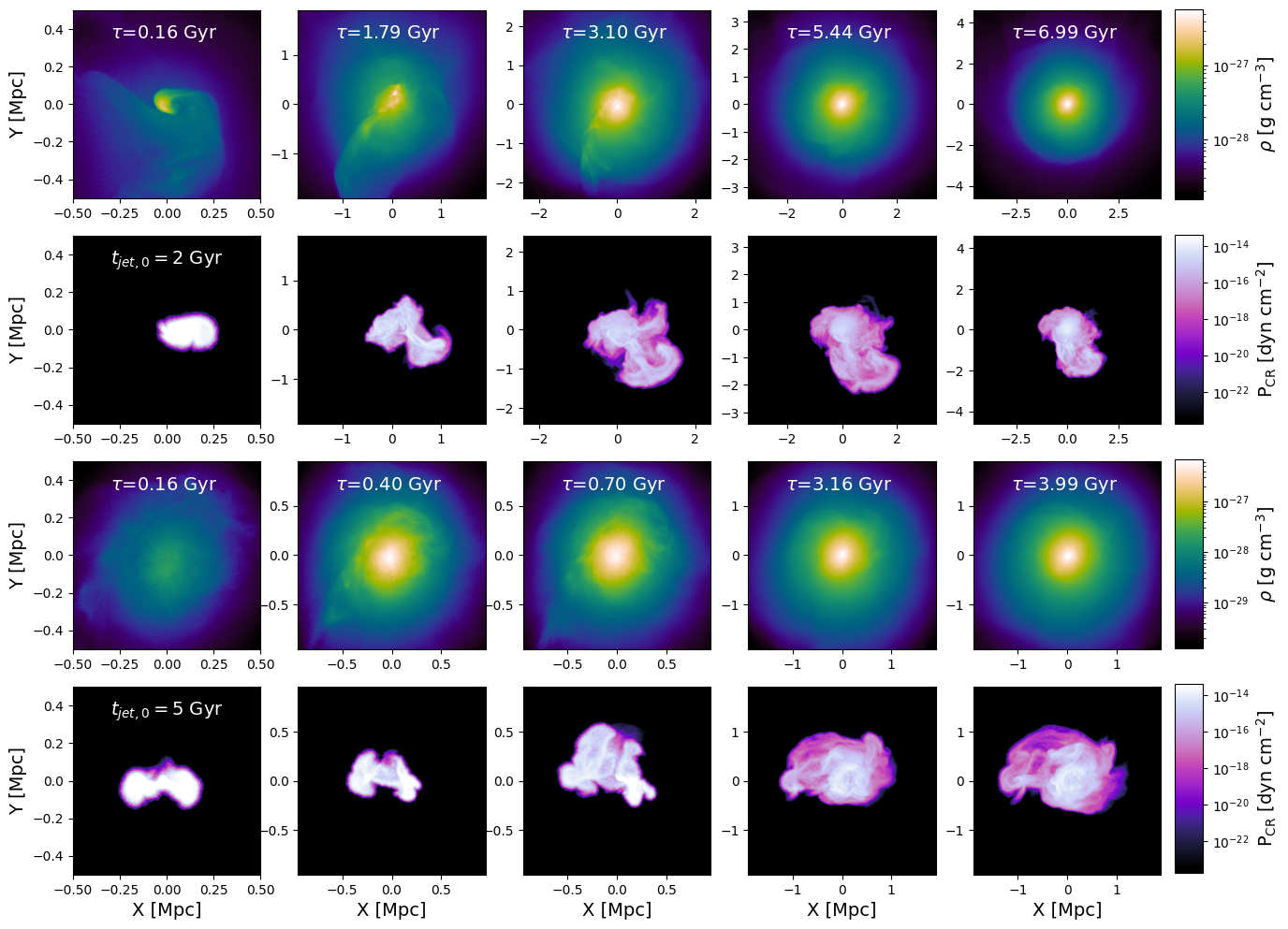}
    \caption{Case $R=1:2$, $\theta=10^{\circ}$ and starting jet activity after 50 Myr and 1 Gyr (upper panels), and 2 Gyr and 5 Gyr (lower panels). Note that Note that $\tau=t - t_{\mathrm{jet},0}$ and that the times shown are different for the 5 Gyr case.}
    \label{fig:maps_r0p5_theta10}
\end{figure*}

\begin{figure*}
    \centering

    \includegraphics[width=0.85\textwidth]{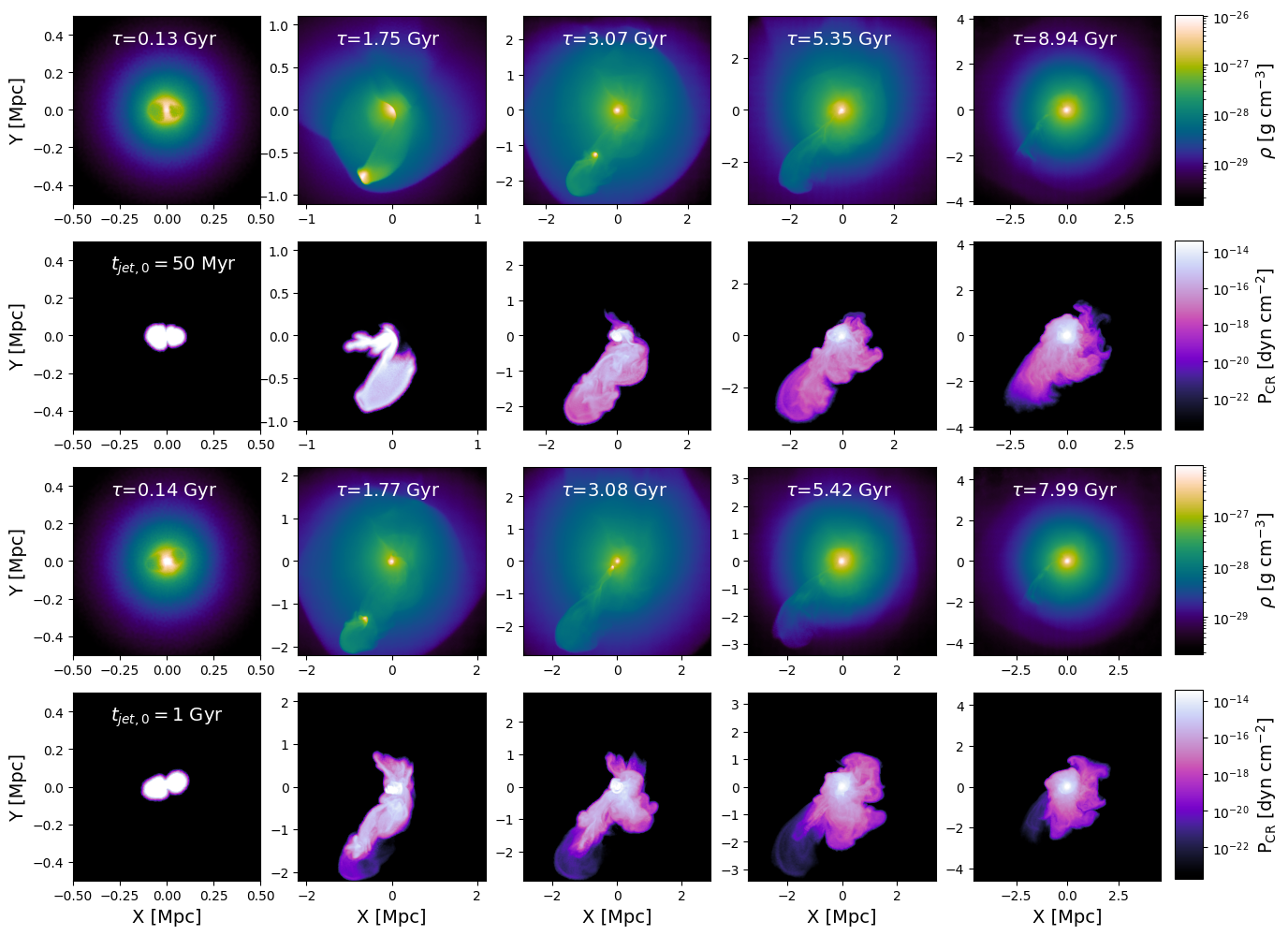}
    \includegraphics[width=0.85\textwidth]{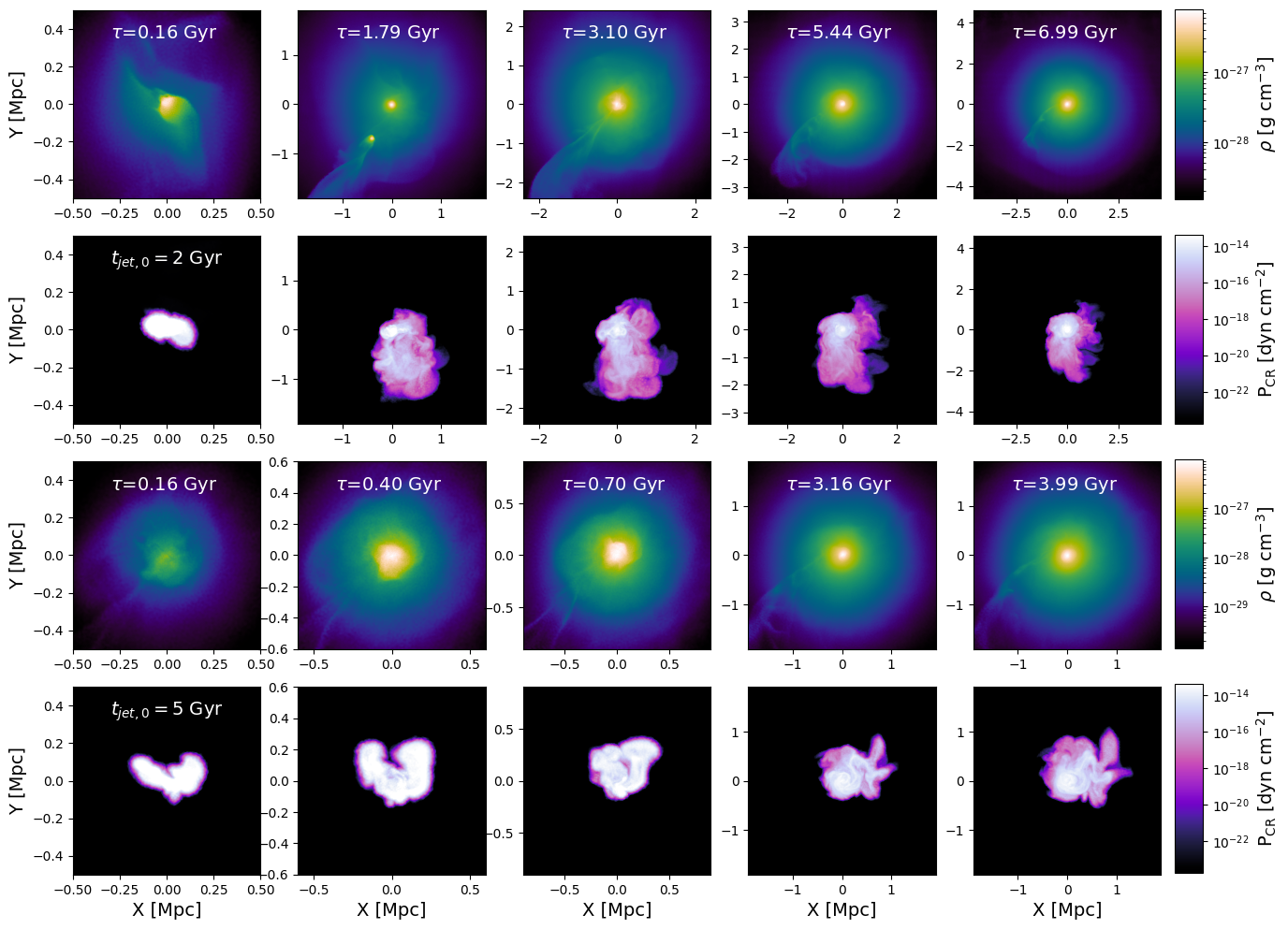}
    \caption{Case $R=1:5$, $\theta=10^{\circ}$ and starting jet activity after 50 Myr and 1 Gyr (upper panels), and 2 Gyr and 5 Gyr (lower panels). Note that Note that $\tau=t - t_{\mathrm{jet},0}$ and that the times shown are different for the 5 Gyr case.}
    \label{fig:maps_r0p2_theta10}
\end{figure*}

We show the projection maps for all the $\theta=20^{\circ}$ and $\theta=10^{\circ}$ runs. For $\theta=20^{\circ}$ runs, we show the projection maps with a mass ratio $R=1:2$ in Fig.~\ref{fig:maps_r0p5}, and with a mass ratio $R=1:5$ in Fig.~\ref{fig:maps_r0p2}. Similarly, for $\theta=10^{\circ}$ runs, we show the projection maps with a mass ratio $R=1:2$ in 
Fig.~\ref{fig:maps_r0p5_theta10}, and with a mass ratio $R=1:5$ in Fig.~\ref{fig:maps_r0p2_theta10}.
In each figure, we show density projection maps and CR pressure maps for the simulation runs where the jet direction was initialized in the $x$-direction. 

%
%
%
%
%
%

\begin{figure*}
    \centering

    \includegraphics[width=0.91\columnwidth]{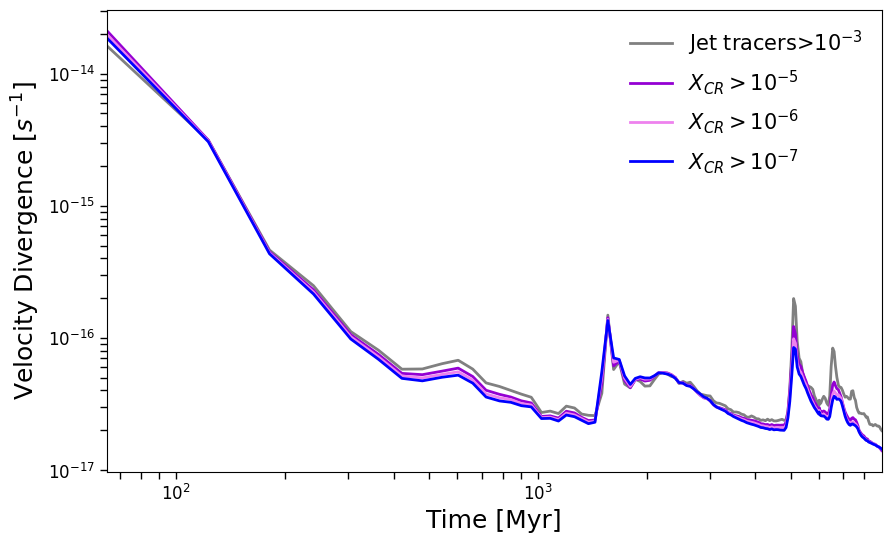} 
    \includegraphics[width=0.89\columnwidth]{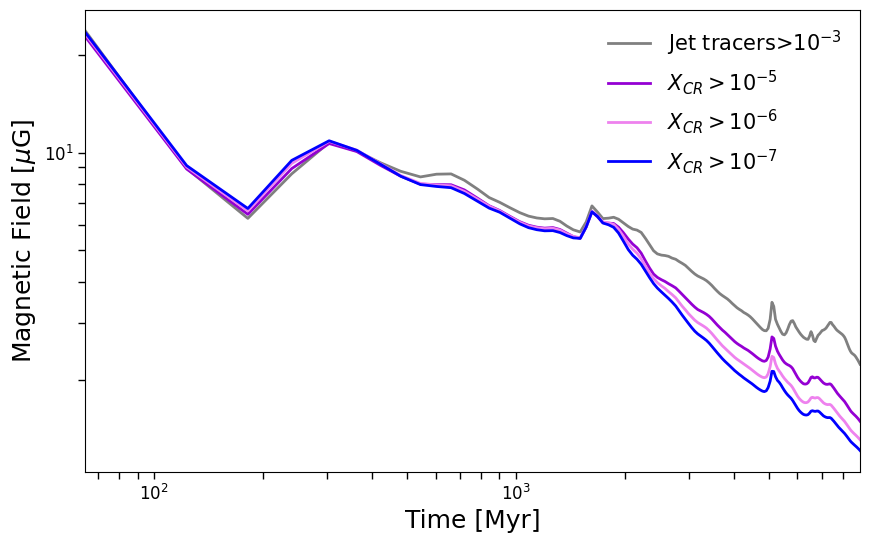}
    \caption{Volume averaged velocity divergence (left) and magnetic field strength (right) of the CR jet material as a function of time for the case $R=1:5$, $\theta=20^{\circ}$ and $t_{\mathrm{jet,0}}=50$ Myr. We show the corresponding evolution for different thresholds described in Appendix~\ref{app:jet_material}.}
    \label{fig:tracers}
\end{figure*}
\begin{figure*}
    \centering

    \includegraphics[width=0.85\textwidth]{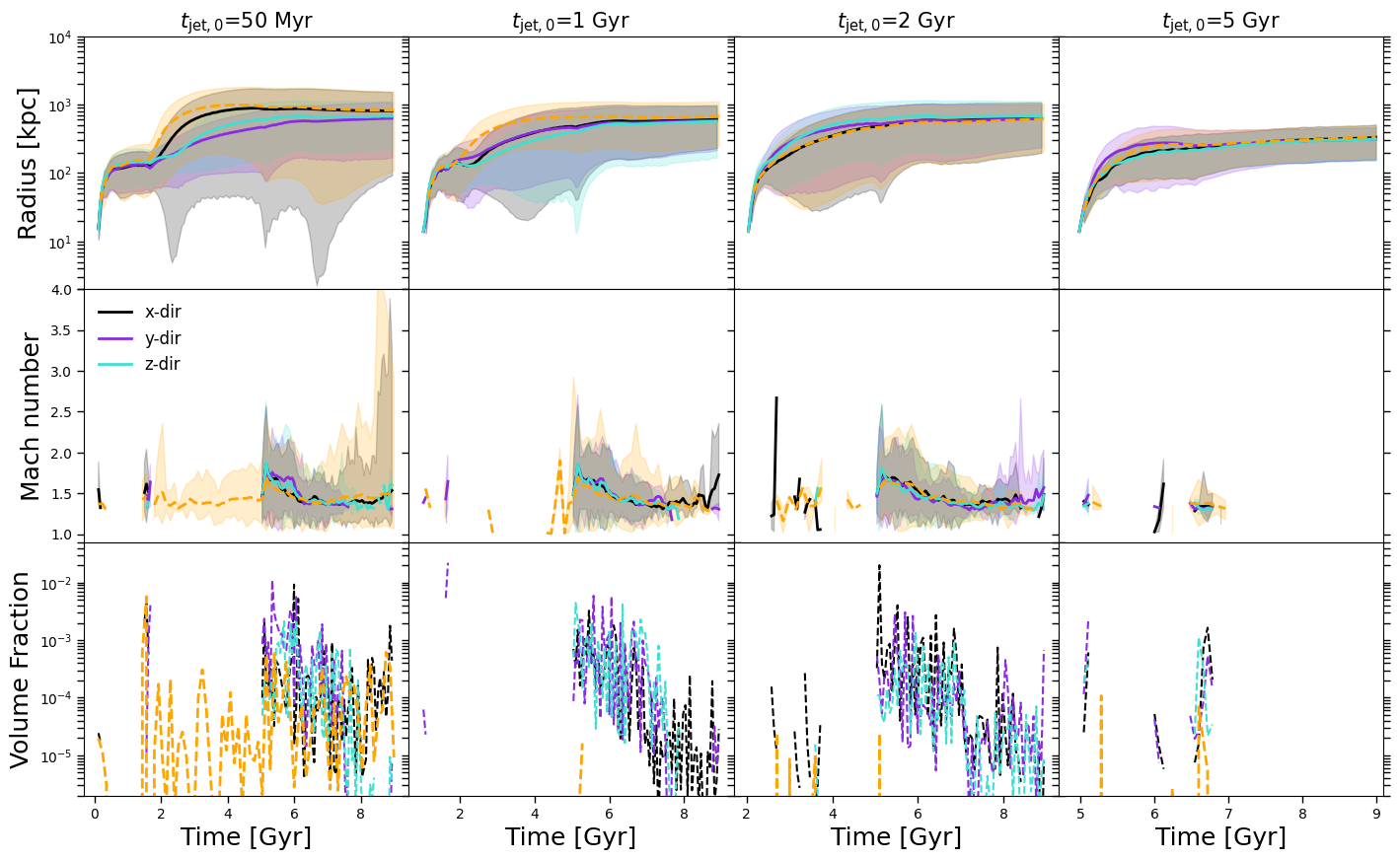}
    \caption{Same as Fig.~\ref{fig:multipanel_evolution} for the $R=1:5$ cases. We overplot in orange the results of the higher resolution simulations. 
    }
    \label{fig:multipanel_evolution_appendix}
\end{figure*}
%

\section{Tracing the jet material}
\label{app:jet_material}

%
We monitor jet material by setting a threshold for the cosmic ray pressure to thermal pressure ratio, $\chi_{\mathrm{CR}}=P_{\mathrm{CR}}/P_{\mathrm{th}}$. An additional method involves using the already defined AREPO scalar jet-tracer output. In Fig.~\ref{fig:tracers}, we present the volume-averaged evolution of velocity divergence and magnetic field strength in the jet material, using various thresholds. Throughout this work, we used $\chi_{\mathrm{CR}}=10^{-7}$. While selecting different thresholds could introduce slight variations in the statistics, as illustrated in Fig.~\ref{fig:tracers}, these differences are minimal and do not significantly impact our main results. More attention on the CR pressure ratio, $\chi_{\mathrm{CR}}$, would only be necessary in the case in future work for the modelling of synchrotron emission. For the volume averaged radius shown in Figs.~\ref{fig:multipanel_evolution} and ~\ref{fig:multipanel_evolution_theta10}, we compute the radii at each of the desired cells knowing the cell coordinates and considering the coordinates of the black-hole as the center. Finally, having all the information of the radii at the cells where the CR jet material is, we take a volume average.

\section{Higher resolution runs}
\label{app:high-res}

In this section, we show the time evolution of the jet material in our high-resolution runs (see Table~\ref{table:sims}). In Fig.~\ref{fig:multipanel_evolution_appendix} we show the same time evolution as shown in Fig.~\ref{fig:multipanel_evolution} for the $R=1:5$, $\theta=20^{\circ}$ runs. We show the results of the high-resolution runs in orange color. The high-resolution runs are limited to the cases where we set the jet to ignite in the $x$-direction. In the first row, we see how the CR material reaches the same radius as the low-resolution runs. In particular, we see that at these two different resolutions, the CR material in these $R=1:5$ cases reaches around 2 Mpc by the end of the simulation. In the second row, we see that at higher resolution, the shock finder tends to pick-up more shocks, which is expected. The majority of the newly detected shocks at higher resolution are rather weak, with $\mathcal{M}\lesssim 1.4$. In addition, the volume fraction of CR material that encounters these newly detected shocks amounts to $\lesssim 1$\%. These shocks are not expected to be relevant in the full DSA regime \citep[see, e.g.,][]{2019ApJ...876...79K,2021ApJ...915...18H}. 
Therefore, for the purposes of this study, we can safely disregard these weak shocks, as they are unlikely to be relevant for potential radio emissions associated with radio relics.

We defer a high-resolution study, which would demand more computational resources, to future work. Such higher resolution simulations are more crucial for analyzing the early phase of inflated lobes, where internal shocks and turbulence significantly influence the energy transfer and impact the resulting radio emission. 
%
%

\section{Vortex-p analysis: Solenoidal and compressive components}
\label{app:vortex}

The details of the vortex-p code can be found in \citealt{Valles_Perez_2024}. In this appendix, we only summarize the key steps and parameters used for this work. First, the AREPO velocity field data is interpolated onto a set of nested adaptive mesh refinement (AMR) grids, according to our local resolution in different regions. 
We run the decomposition in our computational box of $L=40$ Mpc. The base grid has a resolution of $N_x = 128$ and a maximum of $n_l = 6$ refinement
levels, yielding a peak resolution
of $\Delta x \approx$ 4.8 kpc. Next, the algorithm 
relies on the Helmholtz-Hodge decomposition of the velocity field that has already been assigned onto the AMR grid. We considered a minimum number of 16 particles within the kernel around a cell. 
The multi-scale filtering algorithm in vortex-p determines the local filtering
scale, $L_0$, in an iterative way searching for convergence on the turbulent velocity field.
We follow \citealt{Valles_Perez_2024} and use a fiducial filter tolerance of 10\%, 5\% increment in the growing step, and a maximum of 200 iterations \citep[see also][Section 2.2]{Valles_Perez_2021b}. To avoid a divergent behavior at discontinuities, 
the iteration stops at shocked cells with a threshold Mach number, $\mathcal{M}_{th} \geq 2$. 
Finally, the solenoidal and compressive velocities are mapped back from the internal AMR grid to the original particle positions. We verified that the algorithm relative errors are of the same order as the errors involved in the initial grid assignment
\citep[see Figs. 2 and 3 in][]{Valles_Perez_2024}.

\begin{figure}
    \centering
    \includegraphics[width=0.8\columnwidth]{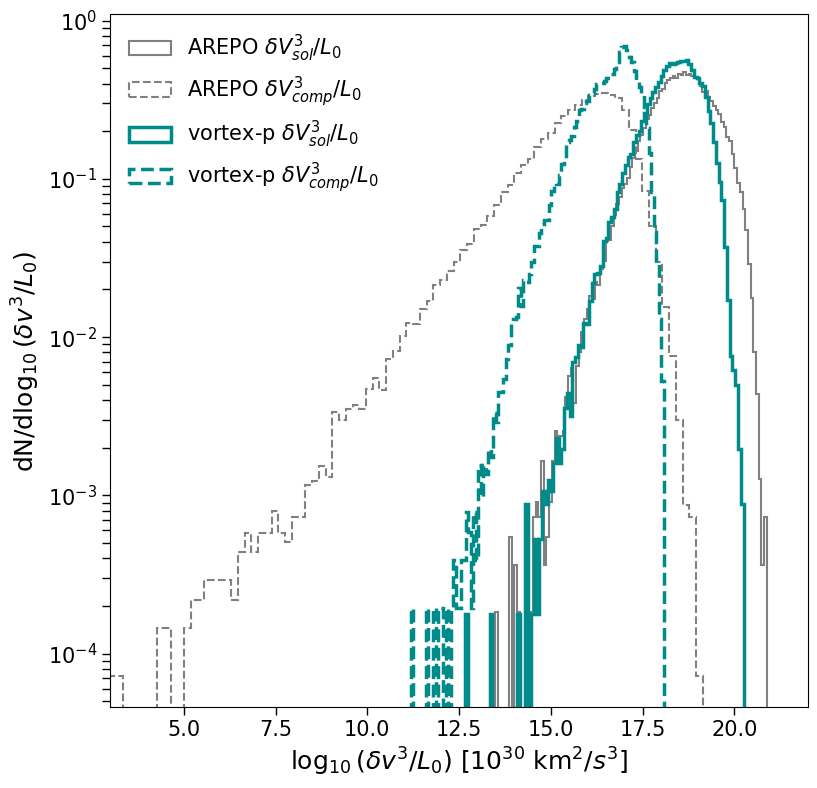}
    \caption{Volume weighted $\delta v^3/L_0$ distributions at 2.9 Gyr for the $R=1:5$,  $\theta=20^{\circ}$ run. We show the distributions corresponding to the solenoidal (solid lines) and compressive (dashed lines) turbulent velocities as computed by the two methods described in Section~\ref{sec:halos}. The first method (taking into account the AREPO output) is shown in gray color and the second method (using the vortex-p code). We multiplied by a factor of $10^{30}$ for purposes of better visualization.}
    \label{fig:vortex-p}
\end{figure}

In Fig.~\ref{fig:vortex-p} we show a histogram of the turbulent energy rate, $\delta v/L_0$, computed with two methods described in Section~\ref{sec:halos} for the $R=1:5$, $\theta=20^{\circ}$ run. We show the distributions corresponding both to the compressive and solenoidal turbulent velocities. This rate is used to calculate the turbulent energy flux in Eq.~\ref{eq:flux}. As can be seen in Fig.~\ref{fig:vortex-p}, the turbulent energy rate from vortex-p is in good agreement with the turbulent energy rate computed with the first method. This implies that assuming the solenoidal turbulence in the simulation is in the inertial range down to the resolution scales is a reasonable proxy for computing $F_{\mathrm{turb}}$. On the other hand, the distributions of compressive turbulent velocities differ between the two methods. This would be relevant for other turbulent re-acceleration models that have not been considered in this work.\\

\bibliography{paola}{}
\bibliographystyle{aasjournal}



\end{document}